\documentclass[aps,prl,reprint,superscriptaddress,amsmath,amssymb]{revtex4-2}
\usepackage{graphicx}
\usepackage{amsmath}
\usepackage{color}
\usepackage{bm}
\usepackage{braket}
\usepackage[export]{adjustbox}
\usepackage[colorlinks=true,allcolors=purple]{hyperref}
\usepackage{dcolumn}
\usepackage{bm}
\usepackage{tikz}
\usepackage{titlesec}
\usepackage[caption=false]{subfig}

\usepackage{pdfpages}
\usepackage{pgffor}

\makeatletter
\AtBeginDocument{\let\LS@rot\@undefined}
\makeatother

\begin{document}
\newenvironment{captivy}[1]{
  \begin{tikzpicture}[every node/.style={inner sep=0}]
    \node[anchor=south west,inner sep=0] (image) at (0,0) {#1};
    \begin{scope}[x={(image.south east)},y={(image.north west)}]
}
{
        \end{scope}
  \pgfresetboundingbox
  \path[use as bounding box] (image.south west) rectangle (image.north east);
  \end{tikzpicture}%
}
\newcommand*{\oversubcaption}[3]{
  \draw (#1) node[fill=white,inner sep=0pt, opacity=0.2, above, yscale=1.1, xscale=1.1] {\phantom{(a)#2}};
  \draw (#1) node[inner sep=0pt, above]{%
    \subfloat[#2\label{#3}]{\phantom{(a)}}
  };
}
\preprint{APS/123-QED}
\title{Superfluidity in active quantum flocks }
\def\afflux{ Department of Physics and Materials Science,\\ University of Luxembourg, L-1511 Luxembourg, Luxembourg}
\def\affaugs{ Theoretical Physics III, Center for Electronic Correlations and Magnetism,\\
Institute of Physics, University of Augsburg, D-86135 Augsburg, Germany}
\author{Byjesh N. Radhakrishnan }%
\affiliation{\afflux}

\author{Reyhaneh Khasseh }%
\affiliation{\affaugs
}
\author{Thomas Schmidt}%
\affiliation{\afflux}
\author{Markus Heyl} 
\affiliation{%
\affaugs}
\affiliation{Centre for Advanced Analytics and Predictive Sciences (CAAPS), University of Augsburg, Universitätsstr. 12a, 86159 Augsburg, Germany}
\date{\today}
\begin{abstract}
Active quantum matter has very recently emerged at the intersection between bio- and quantum many-body physics, combining the self-organization of living systems with the coherence of the quantum world. Active quantum systems have been shown to exhibit flocking — a collective phenomenon with no precedent in equilibrium quantum physics. In this work we uncover an unexpected layer of quantum order in active quantum flocks: they can become superfluid. We show that, in addition to the symmetry breaking associated with their directed motion, these flocks can also break an additional U(1) symmetry, giving rise to off-diagonal long-range order. For a microscopic model of active hard-core bosons governed by Lindblad dynamics, we derive an effective long-wavelength description of the single-particle density matrix and demonstrate that the flocking phase develops an instability toward off-diagonal long-range order characteristic of superfluid behavior. Our findings reveal active quantum matter as a promising research direction for discovering exotic nonequilibrium phases of quantum matter.
\end{abstract}

\maketitle

\paragraph{Introduction---}
Flocking — the self-organized, aligned motion seen in schools of fish, flocks of birds, and synthetic active matter — is one of the most striking archetypes of collective behavior in nature \cite{Visceck_1995,TONER_2005,Chate_2020}. Starting from the seminal Vicsek model \cite{Visceck_1995}, classical flocking has been studied from a multitude of different angles \cite{Shaebani_2020,Chate_2020,Bricard_2013,Geyer_2018,Kaiser_2017}. The quantum counterpart to flocking, by contrast, has remained largely elusive, as in general the quantum analogue of active matter poses fundamental conceptual questions. What does "activity" even mean in the quantum world? Very recently, this question has begun to find first answers from several directions: proposals for active motion in quantum systems \cite{Antonov_2025,penner_2025,Nadonly_2025,adachi_2022,Taskan_2024,yuan_2024,steiner2026,burgardt2026,Antonov_2026} and the discovery of active quantum flocks \cite{khasseh_2024} — dissipative quantum many-body systems that exhibit flocking alongside genuine quantum features such as a long-distance quantum coherence. Yet a fundamental question remains: beyond these first steps, what truly collective quantum behaviors can active quantum matter sustain? What novel forms of order, with no classical counterpart, can emerge when self-propulsion and alignment meet quantum properties?

In this Letter, we provide evidence that active quantum flocks can exhibit superfluidity, spontaneously breaking a U(1) symmetry in addition to the symmetry associated with their directed motion. To establish this, we develop an analytical hydrodynamic theory for the single-particle density matrix. Starting from the microscopic Lindblad dynamics of hard-core bosons with self-propulsion and alignment interactions, we derive closed equations of motion for the two-point correlation functions. By analyzing the steady-state solutions in the flocking phase, we identify a distinct dynamical instability for the off-diagonal elements of the single-particle density matrix. This instability provides strong evidence for the spontaneous formation of off-diagonal long-range order — the defining signature of superfluidity \cite{Penrose_1956,Kiely_2022,Giamarchi_2003,Haldane_1981}. Our results thus unveil a novel nonequilibrium phase in which macroscopic directed motion coexists with macroscopic quantum coherence, with no classical analogue.

\paragraph{Quantum flocking and emergent long-range coherence---}
We consider an active quantum flocking model composed of two species of hard-core bosons, labeled by an effective spin $\sigma = \pm$, on a one-dimensional periodic lattice of length $L$. The key ingredients of flocking, self-propulsion and alignment interactions~\cite{Visceck_1995}, are implemented via dissipative processes. The resulting dynamics is governed by a Lindblad master equation~\cite{gardiner_2004}:
\begin{equation}
\begin{split}
       \frac{d \hat{\rho}}{dt}
    &=
    -i\left [\hat{H},\hat{\rho} \right ] 
    +
    \mathcal{D}_{\mathcal{A}}(\hat{\rho})
    +
    \mathcal{D}_{\mathcal{M}}(\hat{\rho}) ,
    \label{eq:linblad}
\end{split}
\end{equation}
where $\hat{\rho}$ is the many-body density matrix. 
The dynamics in Eq.~\eqref{eq:linblad} incorporate both coherent evolution, governed by the system Hamiltonian $\hat{H}$, and the dissipation due to the environment via the superoperators $\mathcal{D}_{\mathcal{A}}(\hat{\rho})$ and $\mathcal{D}_{\mathcal{M}}(\hat{\rho})$.

The dissipative operators $\mathcal{D}_{\mathcal{M}}(\hat{\rho})$ and $\mathcal{D}_{\mathcal{A}}(\hat{\rho})$ implement, respectively, self-propulsion and alignment interactions among the spins. In general, the dissipator $\mathcal{D}_{X}(\hat{\rho})$ associated with a process takes the standard Lindblad form,
$
       \mathcal{D}_{X}(\hat{\rho})
       =
       \sum_{l=1}^{L} \sum_{\sigma=\pm} \frac{\Gamma _{X}}{2}
       \big (  2 \hat{X}_{l\sigma} \hat{\rho } \hat{X}_{l\sigma}^{\dagger}
       -
       \hat{X}_{l\sigma}^{\dagger} \hat{X}_{l\sigma} \hat{\rho } 
       -
       \hat{\rho } \hat{X}_{l\sigma}^{\dagger} \hat{X}_{l\sigma} \big ),
$
where $\hat{X}_{l\sigma}$ is a quantum jump operator acting on particles with effective spin $\sigma$ at lattice site $l$, and $\Gamma_X$ is the corresponding rate.

Active motion is implemented by directional quantum jump operators of the form
$
\hat{\mathcal{M}}_{l+}=\hat{c}_{l,+}^{\dagger}\hat{c}_{l+1,+}
$
and 
$
\hat{\mathcal{M}}_{l-}=\hat{c}_{l+1,-}^{\dagger}\hat{c}_{l,-}
$,
where $\hat{c}_{l\sigma}^{\dagger}$ creates a hard-core boson with spin $\sigma$ at site $l$. These processes induce biased motion, such that $\sigma=+ (-)$ particles propagate preferentially to the left (right), thereby generating persistent currents characteristic of active matter.

Alignment interactions, which tend to orient the internal degree of freedom $\sigma$ of a particle according to its neighbors, are described by the jump operators
$\hat{\mathcal{A}}_{l \sigma}=\hat{c}_{l\sigma}^{\dagger} \hat{c}_{l\bar{\sigma}} \hat{\mathcal{P}}_{l}$
where 
$\hat{\mathcal{P}}_{l}=1-  (\kappa/ 2r)  \hat{m}_{l} \hat{M}_{l}\label{eq:al}$ and $\bar{\sigma}$ denote the complement of $\sigma$.
The local density and magnetization at the site $l$ are defined as $\hat{\rho}_{l}=\sum_{\sigma}  \hat{c}_{l,\sigma}^{\dagger} \hat{c}_{l,\sigma}$ and  $\hat{m}_{l}=\sum_{\sigma} \sigma \hat{c}_{l,\sigma}^{\dagger} \hat{c}_{l,\sigma}$, respectively, while $\hat{M}_{l}=\sum_{\substack{|j|=1}}^{r}\hat{m}_{l+j}$ is the surrounding magnetization at the site $l$ within the interaction radius $r$. The operator $\hat{\mathcal{P}}_{l}$ conditions spin flips on the local magnetic environment, favoring configurations with locally aligned spins, and $\kappa$ controls the strength of this alignment.

Coherent dynamics is generated by the Hamiltonian, $\hat{H}=-g\sum_{l=1}^{L}  \sum_{\sigma}\hat{c}_{l,\sigma}^{\dagger} \hat{c}_{l,\bar{\sigma}} $, which induces local spin flips with amplitude $g$, and the current of local spin-flip is defined as $\hat{f}=i\sum_{\sigma} \sigma \hat{c}_{l,\sigma}^{\dagger} \hat{c}_{l,\bar{\sigma}}$. While the Hamiltonian generates quantum superposition, it competes with the dissipative alignment mechanism and therefore tends to suppress flock formation.

The model is invariant under a discrete $\mathit{\mathbb{Z}}_2$ symmetry corresponding to the simultaneous reversal of all particle spins and their direction of motion.
The emergence of collective motion therefore requires spontaneous symmetry breaking, making the global magnetization $\hat{M}=\sum_{l=1}^{L}\hat{m}_{l}$ the suitable order parameter. The interplay of directional motion and dissipative alignment drives the system intrinsically out of equilibrium and violates local detailed balance~\cite{khasseh_2024}. Above a critical alignment strength $\kappa>\kappa_c$, the system enters a ferromagnetic flocking phase characterized by a macroscopic magnetization and collective transport in a spin-selected direction, realizing a genuinely quantum analogue of classical flocking in one-dimension.

The quantum nature of this phase is revealed by the emergence of long-range coherence, which can be quantified by
$
C(t)=\sum_{i,j=1}^{L}\sum_{ \nu _{ij} \ne \nu _{i^{'}j^{'}}} \left | \langle \nu_{ij}|\hat{\rho}_{c}(t)|\nu _{i^{'}j^{'}} \rangle  \right |,
$
where the operator $\hat{\rho}_{c}(t)$ is the reduced density matrix of two sites $i$ and $j$, obtained after tracing out all other degrees of freedom, and $| \nu _{ij}\rangle=| \mathbf{n} _{i},\mathbf{n}_{j}\rangle $ are the local basis states with $\mathbf{n} _{i} \in \{ \o ,+,- ,\pm \}$. Numerical simulations show that, in the flocking phase, these off-diagonal matrix elements remain finite even at large separations, signaling long-range quantum coherence, while they vanish in the disordered phase~\cite{khasseh_2024}. Collective motion in quantum flocks is thus intrinsically accompanied by nonclassical correlations encoded in the off-diagonal structure of the density matrix.

One of the contributions to this coherence arises from one-body correlators $ \langle \hat{\Pi}_{ij}^{\sigma_{1} \sigma_{2}} \rangle$, which have the form
\begin{equation}
\begin{split}
  \hat{\Pi}_{ij}^{\sigma_{1}\sigma_{2}} 
  = \hat{c}_
{i \sigma_{1}}^{\dagger}\hat{c}_{j \sigma_{2}} 
\quad \text{with} 
\quad 
i \ne j ,
\end{split}
\end{equation}
where the notation $\langle . \rangle $ denotes the expectation defined as $\text{Tr}\{ \hat{\rho}(t) \hat{\mathcal{O}}  \}$ and $\text{Tr}\{.\}$ denotes the trace of an operator. These correlators characterize phase coherence between distant sites and therefore provide a direct probe of long-range quantum order. Developing a coarse-grained hydrodynamic description for these quantities offers a natural route to understanding the emergence, structure, and dynamics of long-range coherence in the flocking phase.

\begin{figure*}
    \centering
 \includegraphics[clip,trim=0cm 0cm 0cm 0cm,width=.29\textwidth]{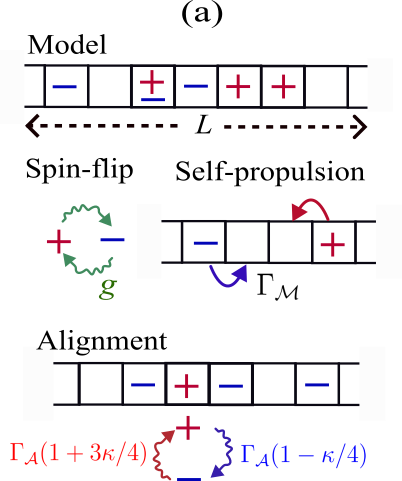}
 \hfill
 \includegraphics[clip,trim=0cm 0cm 0cm 0cm,width=.27\textwidth]{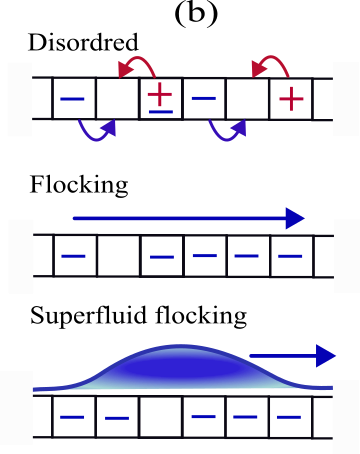}
 \hfill
 \includegraphics[clip,trim=0cm 0cm 0cm 0cm,width=.4\textwidth]{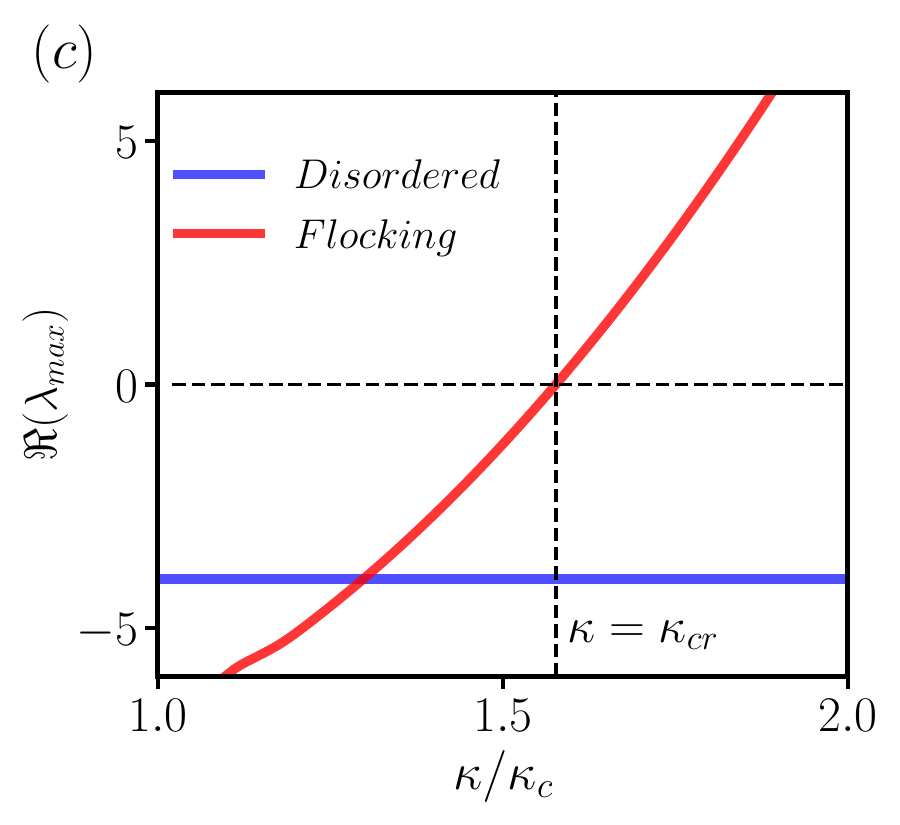}
 \hfill
\caption{(a) The schematics showing model and interactions. The rate of alignment induced flips are calculated by setting $r=2$. (b) Schematic representation of different phases, and (c) the real part of largest eigenvalue ($\lambda_{max}$) of the matrix $(\mathcal{K}_{0}+\mathcal{K}_{m} )$ in the flocking and disordered phase.  The parameters chosen as $\Gamma=2$, $g=3$ (disordered) and $g=0.3$ (flocking). The eigenvalues cross the zero line for $\kappa=\kappa_{cr}$ and shows crossover from uncorrelated to correlated motions of spins. The flocking state is stable only for $\kappa > \kappa_c (1+2(g/\Gamma)^2)$, and hence the $x$-axis start from $\kappa_{c}$.}
    \label{fig:dynamics}
\end{figure*}

\paragraph{Hydrodynamics of two-point correlation functions---}
 Starting from the Lindblad master equation, we derive closed equations of motion for the two-point correlators $ \langle \hat{\Pi}_{ij}^{\sigma_{1} \sigma_{2}}  \rangle$. Their dynamics contains three distinct contributions: (i) coherent evolution generated by the Hamiltonian via local spin flips, (ii) dissipative transport arising from active motion, and (iii) dissipative alignment of the internal spin degree of freedom  [\nameref{sec:endmatter}].

Assuming spatially homogeneous density, magnetization and spin-flip current in the bulk, $\left \langle \hat{\rho}_{i} \right \rangle =\rho(t)$, $\langle \hat{m}_{i}\rangle =m(t)$, $\langle \hat{f}_{i}\rangle =f(t)$, and $\big \langle \hat{M}_{i}\big \rangle =\bar{M}$ for all lattice sites $i$, and performing a controlled factorization of nonlinear operator products, the homogeneous dynamics reduces to a closed linear system \cite{mykey}:
\begin{equation}
    \begin{split}
    \frac{d}{dt}\mathbf{\Pi}
    =
    \big (\mathcal{K}_{0}+\mathcal{K}_{m}  \big ) \cdot \mathbf{\Pi}
    +
    \mathcal{O}(m^{3}) ,
    \label{eq:rho_mat}
    \end{split}
\end{equation}
where $\mathbf{\Pi} =\big (  \langle \hat{\Pi}_{ij}^{++}  \rangle,  \langle \hat{\Pi}_{ij}^{--}  \rangle , \langle \hat{\Pi}_{ij}^{+-}  \rangle,  \langle \hat{\Pi}_{ij}^{-+}  \rangle \big )^{T}$. The matrices $\mathcal{K}_{0} $ and $\mathcal{K}_{m}$ are $4 \times 4$ matrices with coefficients depending on $\rho$, $m$ and $f$, and are controlled by the microscopic parameters $g$, $\kappa$ and $\Gamma $ (hereafter we set $\Gamma_{\mathcal{A}/\mathcal{M}} =\Gamma$).
Explicitly,
\begin{equation}
\mathcal{K}_{0}
=-\frac{1}{2}
\begin{pmatrix}
A & B \\
B  &A \\
\end{pmatrix},
\quad 
\mathcal{K}_{m}
=
\begin{pmatrix}
C & \frac{im}{2}S \\
\frac{im}{2} T &D\\
\end{pmatrix} ,
\label{eq:matrix_form}
\end{equation}
where $A=4\Gamma I$ and $B=ig (\sigma_{x}-I)$ with $I$ denoting the two-dimensional identity matrix and $\sigma_{x,y,z}$ are the Pauli matrices. The diagonal blocks are given by $C=diag \big (c_{+},c_{-}\big )
$ and  $D=diag \big (d_{+},d_{-}\big )$ with
\begin{equation}
\begin{split}
    c_{\pm } = &\mp 
2gf - \Gamma \left [  \mp \kappa m 
 - m^2  \big ( \kappa^{2} +2\kappa  \big )  \right ] ,\\
     d_{\pm}= &
- \Gamma m^2 \big ( \kappa^2  \pm \kappa \big ) ,
\label{eq:c,d}
\end{split}
\end{equation}
while the off-diagonal blocks take the form $S=-2g (\sigma_x -i\sigma_y )-\alpha$ and $T=-2g (I+ \sigma_x )+\alpha$ with $\alpha=\Gamma f \kappa  (I- \sigma_x )/2$.

Equations~\eqref{eq:matrix_form}–\eqref{eq:c,d} incorporate all contributions up to second order in the magnetization, while subleading corrections in the interaction radius $r$ are neglected [\nameref{sec:endmatter}], as the mean-field limit assumes the largest possible interaction range. The leading order dependence on $m$ reflects the ferromagnetic character of the alignment interaction, whereas higher-order terms encode nonlinear feedback between local order and long-range correlations. As a result, the phase-dependent value of $m$ directly controls the dynamical evolution of correlations.

It is important to note that the closure of the hydrodynamic equations requires a consistent factorization of nonlinear operator products. To this end, we employed a Wick-type decoupling scheme retaining all allowed contractions, 
\begin{equation}
    \begin{split}
\big \langle  \hat{n}_{j,\sigma_1} \hat{\Pi}_{ij}^{\sigma_1 \sigma_2} \big \rangle 
&=
   \big \langle \hat{n}_{j,\sigma_1} \big \rangle \big \langle \hat{\Pi}_{ij}^{\sigma_1 \sigma_2} \big \rangle 
 +
 \big \langle \hat{\Pi}_{jj}^{\sigma_1 \sigma_2} \big \rangle  \big \langle \hat{\Pi}_{ij}^{\sigma_1 \sigma_1} \big \rangle ,
    \end{split}
\end{equation}
which retains both mean-field and exchange contributions.  The motivation for this particular factorization scheme is as follows: in the absence of active motion and alignment interactions, the system should exhibit Larmor precession within the subspace containing a single particle on the lattice. While a purely mean-field factorization fails to capture this limit, the Wick-type decoupling does so within the homogeneous approximation, yielding a controlled and physically consistent hydrodynamic description of local quantum fluctuations.

 \paragraph{Disordered phase: decay of correlations---}
In the disordered phase, where the magnetization vanishes ($m=0$), the dynamics of the correlation functions is governed solely by the matrix $\mathcal{K}_{0}$. In this limit, all eigenvalues of $\mathcal{K}_{0}$ have strictly negative real parts [FIG.~\ref{fig:dynamics}], implying that the homogeneous state is dynamically stable. As a consequence, the two-point correlators decay exponentially in time,
\begin{equation}
\big \langle \hat{\Pi}_{ij}^{\sigma_{1} \sigma_{2}} (t)\big \rangle
\sim 
\exp\big (-2 \Gamma t \big ) ,
\end{equation}
demonstrating the absence of long-range correlations in the disordered phase. The absence of long-range correlations demonstrates that active motion alone cannot sustain correlated collective behavior, consistent with the numerically established absence of long-range coherence reported in Ref.~\cite{khasseh_2024}. Owing to the simple block structure of $\mathcal{K}_{0}$, its spectrum can be obtained analytically in the large-interaction-radius limit ($r\sim L/2$) within mean-field theory [\nameref{sec:endmatter}]. The resulting exact solution confirms the purely dissipative decay of correlations in the absence of spontaneous symmetry breaking.

 \paragraph{Flocking phase: emergence of coherent dynamics---}
In the flocking phase, the magnetization reaches a finite steady-state value, $m\neq 0$, which qualitatively modifies the dynamics of the correlation functions. To characterize their steady-state behaviour, we evaluate Eq.~\eqref{eq:rho_mat} using the corresponding stationary values of $m$ and $f$. These quantities are obtained by deriving the coarse-grained dynamics of the magnetization and spin-flip current, following the same procedure as for the correlators and assuming spatial homogeneity. Importantly, this calculation also requires inclusion of fluctuations in higher moments of $\hat{m}$, with $\langle \hat{m}^{2} \rangle = m^{2} + \sigma_{m}^{2}$ and $\langle \hat{m}^{3} \rangle = m^{3} + q m$, where $\sigma_{m}^{2}$ and $q$ encode fluctuation corrections determined by microscopic details. The steady-state values of $m$ and $f$ in the homogeneously magnetized state are given as~\cite{khasseh_2024,mykey}
\begin{equation}
    \begin{split}
        m^{2}\approx \frac{1}{2\kappa_{c}^{2}}\frac{\kappa/\kappa_{c} -1-\Delta_{g}}{q-1 / \kappa_c},
        \quad 
         f \approx -\frac{2gm}{ \Gamma} ,
    \end{split}
\end{equation}
where the parameter $\kappa_c = 1/(2\sigma_{m}^{2})$ is determined by the magnetization variance. The term $\Delta_{g}=2g^{2}/\Gamma^{2}$ describes the suppression of magnetic order due to coherent spin flips. The stable flocking phase exists only for $\kappa>\kappa_{c}(1+\Delta_{g})$ and $q>1/\kappa_{c}$ \cite{mykey}. Consistent with Ref.~\cite{khasseh_2024}, the critical alignment strength is renormalized according to 
\begin{equation}
\kappa_{c}(g)=(1+\Delta_{g})\kappa_{c}(0)
\label{eq:kc}
\end{equation}
with $\kappa_{c}(0)\approx2.4$ obtained numerically. While $\kappa_{c}$ follows from the variance of magnetization, the coefficient $q$ depends on microscopic details and must be extracted from simulations. 

The steady-state solutions of Eq.~\eqref{eq:rho_mat} are determined by the eigenvalues of the full dynamical matrix $(\mathcal{K}_{0}+\mathcal{K}_{m})$. As shown in FIG.~\ref{fig:dynamics}, the real part of the largest eigenvalue $(\lambda_{max})$ increases with alignment strength and vanishes at a critical value $\kappa_{cr}$. Since the linearly stable flocking state exists only for $\kappa>\kappa_{c}(1+\Delta_{g})$, the spectrum is analyzed in this regime. At $\kappa=\kappa_{cr}$, the remaining eigenvalues retain negative real parts and their associated modes decay exponentially, leaving a single oscillatory mode that governs the long-time dynamics. This identifies the onset of persistent coherent correlations in the flocking phase.

For sufficiently strong alignment, $\kappa > \kappa_{cr}$, the present theory predicts an unbounded growth of the correlation functions, signaling the breakdown of the two-point closure approximation. A quantitative description of this regime requires incorporating higher-order contributions, which are beyond the scope of this work. Nevertheless, the present hydrodynamic theory captures the emergence of sustained two-point coherence and identifies the parameter window where quantum flocks exhibit robust long-range correlations. In particular, it provides a controlled estimate of the onset and lower bound of steady-state quantum coherence in the flocking phase.

\begin{figure}
    \centering
 \includegraphics[clip,trim=0cm 0cm 0cm 0cm,width=.4\textwidth]{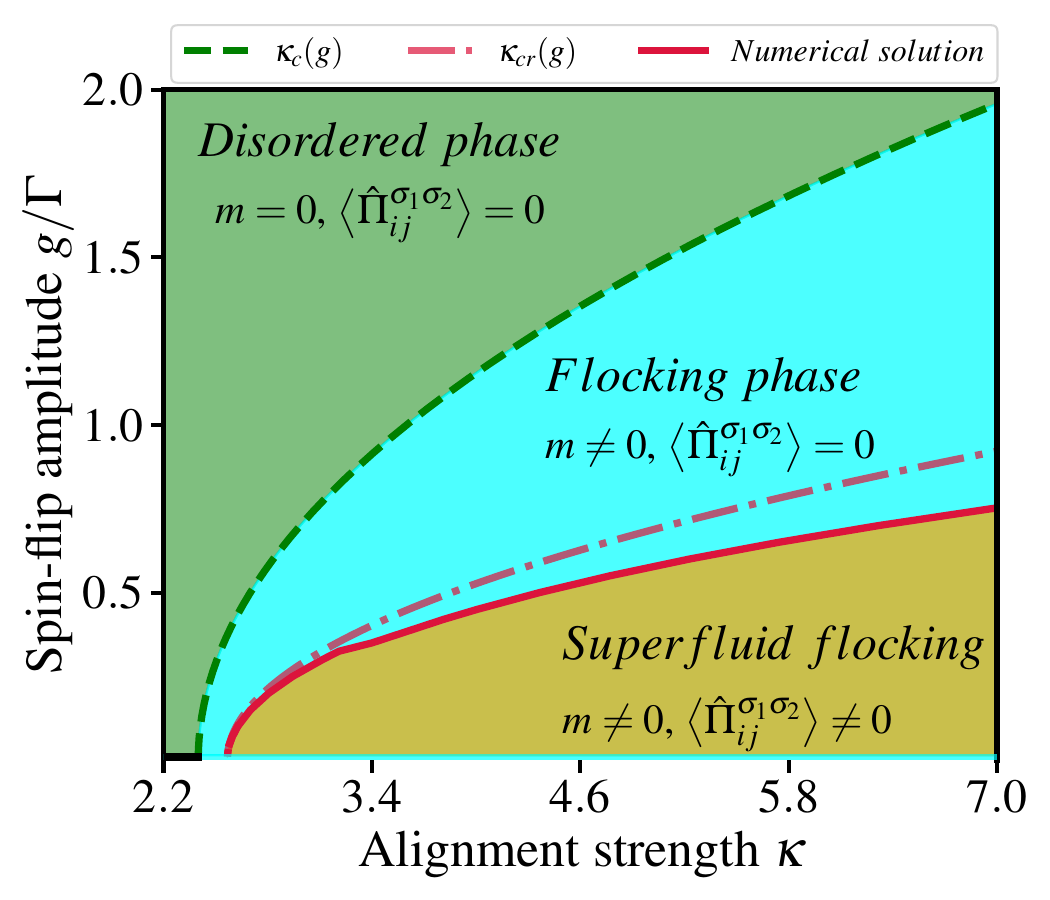
 }
  \hspace{.25cm}
\caption{Phase diagram in the $(\kappa,g)$ plane. The green dashed line, $\kappa_{c}(g)$, is the phase boundary between disordered and flocking states while the red solid line separates uncorrelated and correlated motions in the system. The red dash-dot line, $\kappa_{cr}(g)$, represent the analytical prediction for the same boundary in the small $g$ limit. } 
    \label{fig:phase_diagram}
\end{figure}

 \paragraph{Phase diagram: disordered, flocking and superfluid flocking---}
 The phase diagram in the $(\kappa,g)$ plane follows directly from the dynamics of the correlation functions and characterizes both the onset of collective motion and the emergence of long-range quantum coherence. We first adopt the phase boundary obtained in Ref.~\cite{khasseh_2024}, which separates the disordered and flocking phases. In the presence of coherent spin flips, the critical alignment strength is shifted according to Eq. \eqref{eq:kc} shown as the green dashed line in FIG.~\ref{fig:phase_diagram}. The region above this line corresponds to the disordered phase, while the region below supports collective flocking.

To identify the onset of correlated motion within the flocking phase, we determine the critical interaction strength $\kappa_{cr}(g)$ at which the leading eigenvalue of the correlation matrix becomes purely imaginary. This condition marks the transition from purely decaying to persistent oscillatory correlations and is shown as the red solid line in FIG.~\ref{fig:phase_diagram}. The resulting phase diagram exhibits three distinct regimes. At large $g$, the system is disordered. For intermediate parameters, the system forms a flocking state in which correlations decay at long times, indicating the absence of long-range correlations. Finally, for sufficiently strong alignment and weak spin flips, the system enters a coherent flocking regime characterized by persistent quantum correlations.

In the weak-$g$ limit, the coherence boundary admits the analytical expansion
\begin{equation}
    \kappa_{cr}(g) \approx \kappa_{cr}^{0} \left ( 1+ \frac{8}{\Gamma^2} g^2 + 
     \frac{2  m_{0} \kappa_{cr}^{0}}{\Gamma^4} 
  g^4\right ) ,
\end{equation}
where
$\kappa_{cr}^0(q_0)=\frac{1}{3} \left ( 1+ \gamma^{-1/3}(q_0)+ \gamma^{1/3}(q_0)\right ) \kappa_{c}(0)$ with $\gamma(q_0) = (1+ 108 q_{0} + 6 \sqrt{6} \sqrt{q_{0} (1+54q_{0})})$ and $q_0 =q-1/\kappa_{c}$, and $m_0 $ denotes the magnetization when $g=0$. This analytical prediction, shown as the red dash-dot line in FIG.~\ref{fig:phase_diagram}, is in good agreement with the numerical results.

The lower region of the phase diagram is of particular interest, corresponding to the regime of \textit{superfluid flocks}, where the steady-state one-body correlators remain finite and signal long-range phase coherence. In this regime, collective motion is accompanied by genuine quantum correlations, indicating a superfluid character of the ordered state \cite{Penrose_1956}.

To elucidate the origin of this behavior, we analyze the eigenvector associated with the largest eigenvalue, $\lambda_{\max}$, denoted by $v_{\lambda_{\max}}$. In the symmetry-broken regime where the majority of particles occupy the $\sigma=+$ spin state $(m>0)$, numerical results suggest the ansatz
$
v_{\lambda_{\max}}=
\begin{pmatrix}
x, 0 ,0
\end{pmatrix}^{T}$, with 
$
x=
\begin{pmatrix}
1,
\varepsilon_0
\end{pmatrix}^{T}$.
The parameter $\varepsilon_0$ quantifies the mixing between the correlation functions $\langle \hat{\Pi}_{ij}^{++}\rangle$ and $\langle \hat{\Pi}_{ij}^{--}\rangle$. Near the critical point, where $\Re(\lambda_{\max})\ll 1$, and in the weak spin-flip regime, one finds $\varepsilon_0 \sim g^2/(8\Gamma^2)$ \cite{mykey}. For the parameters considered in FIG. \ref{fig:dynamics}, $\varepsilon_0\approx 10^{-3}\ll 1$, such that the dominant eigenmode is strongly concentrated in a single polarization sector. Consequently, beyond the critical point, $\langle \hat{\Pi}_{ij}^{++}\rangle$ acquires an exponentially growing component, signaling the emergence of a polarized superfluid state. The numerical and analytical calculations of eigenvectors are included in Ref.~\cite{mykey}.

The opposite scenario occurs for $m<0$, where the majority of particles occupy the $\sigma=-$ spin state. In this case, the dominant eigenvector is well approximated by
$
x=
\begin{pmatrix}
\varepsilon_0,
1
\end{pmatrix}^{T}$,
and the exponentially growing mode is transferred to $\langle \hat{\Pi}_{ij}^{--}\rangle$. Thus, the spontaneous breaking of spin symmetry selects the polarization sector that becomes unstable and thereby determines the polarized superfluid state that emerges.
This behavior has no classical counterpart: in the singular limit $g\to0$, quantum correlations vanish, and the system reverts to a classical flock.

While our analysis suggests a clear instability towards off-diagonal long-range order in the considered system, it is important to note that a very recent work has pointed out that the model we consider exhibits a very special structure~\cite{Brunner2026}.
At first sight this might appear as if this would exclude the superfluid instability we study here.
This, however, is not the case.
While for the factorized solution of the steady state for this particular model, the steady state doesn't realize a superfluid, the key implication is that any infinitesimal perturbation yielding a nonzero $\Pi_{ij}^{\sigma_1\sigma_2}$ in the steady state would lead to the exponential buildup of off-diagonal long-range order according to our analysis.
One such infinitesimal perturbation yielding $\Pi_{ij}^{\sigma_1\sigma_2}\not=0$ would be to add to the Hamiltonian $H$ a simple coherent tight-binding hopping term.

The intermediate regime corresponds to flocking without uniform superfluid order, a nonequilibrium phase in which macroscopic collective motion coexists with decaying quantum correlations. In this regime, strong spin-flip processes suppress interparticle coherence and prevent the establishment of long-range phase order, while alignment remains sufficient to sustain directional motion. The resulting state is therefore dynamically ordered but quantum mechanically incoherent. Since our analysis is restricted to the zero-momentum sector, the observed superfluid phase corresponds to a uniform polarized condensate. Possible finite-momentum superfluid instabilities lie beyond the scope of the present work.

The quantitative position of the coherence boundary depends on the parameter $q$, which characterizes the third-order cumulant of magnetization fluctuations and cannot be fixed within mean-field theory. Its lower bound is given by $q=1/\kappa_{c}$, where the correlated and uncorrelated flocking boundaries merge at $g=0$. Increasing $q$ shifts the boundary $\kappa_{cr}(g)$ toward larger $\kappa$ without altering the qualitative topology of the phase diagram. This shift is clear from the analytical expression of the $\kappa_{cr}^0(q_0)$ mentioned previously. 


 \paragraph{Conclusions and outlook---}
\label{sec:conclusion}
 In this work, we developed an analytical framework to investigate the emergence of long-range quantum correlations in an active quantum flocking system. Starting from the microscopic model of dissipative hard-core bosons with self-propulsion and alignment interactions, we systematically derived coarse-grained evolutions for the two-point correlation functions, $\big \langle \hat{c}_{l,\sigma_1} ^{\dagger} \hat{c}_{l \sigma_2}\big \rangle$. This approach reveals a direct connection between alignment-induced collective motion and the buildup of quantum correlations. In particular, it demonstrates that the flocking phase can support genuinely quantum correlated dynamics with superfluid character.

Within the homogeneous approximation, we showed that the one-body correlations responsible for quantum coherence decay rapidly in the disordered phase but remain finite in the flocking phase, consistent with large-scale numerical simulations reported in Ref.~\cite{khasseh_2024}. Our analysis identified a critical alignment strength, $\kappa_{cr}$, above which long-range quantum correlations emerge. Since these correlators correspond to the off-diagonal elements of the one-body density matrix, their persistence signals off-diagonal long-range order, a defining feature of superfluidity. 

The dominant eigenvector analysis of the linearized dynamics reveals that the leading instability is primarily confined to a single polarization sector, with weak admixture from the opposite spin channel. This structure directly explains the emergence of polarized superfluid order beyond the critical point, while clarifying how spin symmetry breaking selects the dynamically unstable sector.
Based on these results, we constructed a phase diagram displaying three distinct nonequilibrium steady states: a disordered phase, a flocking phase, and a superfluid flocking phase. The latter arises at strong alignment and weak spin-flip amplitudes, where collective motion and quantum coherence coexist.

Direct numerical verification of our predictions remains challenging because of the exponential complexity of open quantum many-body dynamics. Nevertheless, rapid progress in numerical and cold-atom platforms is expected to enable experimental tests in the near future. Our results provide a microscopic foundation for superfluid flocking, identify robust quantum signatures of active matter, and suggest that the experimental setup proposed in Ref.~\cite{khasseh_2024} offers a promising route to realizing and probing superfluidity in active quantum systems. More broadly, our framework opens new avenues for exploring genuinely quantum features of nonequilibrium phases and uncovering emergent collective phenomena unique to active quantum systems.

\par
\section*{Acknowledgments }
\noindent We acknowledge various important discussions on active matter systems with Christoph A. Weber and Ricard Alert.
B.N.R. acknowledges support from the FNR grant PRIDE19/14063202/ACTIVE.

\bibliography{apssamp.bib}

@article{Visceck_1995,
  title = {Novel Type of Phase Transition in a System of Self-Driven Particles},
  author = {Vicsek, Tam\'as and Czir\'ok, Andr\'as and Ben-Jacob, Eshel and Cohen, Inon and Shochet, Ofer},
  journal = {Phys. Rev. Lett.},
  volume = {75},
  issue = {6},
  pages = {1226--1229},
  numpages = {0},
  year = {1995},
  month = {Aug},
  publisher = {American Physical Society},
  doi = {10.1103/PhysRevLett.75.1226},
  url = {https://link.aps.org/doi/10.1103/PhysRevLett.75.1226}
}

@article{TONER_2005,
title = {Hydrodynamics and phases of flocks},
journal = {Annals of Physics},
volume = {318},
number = {1},
pages = {170-244},
year = {2005},
note = {Special Issue},
issn = {0003-4916},
doi = {https://doi.org/10.1016/j.aop.2005.04.011},
url = {https://www.sciencedirect.com/science/article/pii/S0003491605000540},
author = {John Toner and Yuhai Tu and Sriram Ramaswamy}
}

@article{Chate_2020,
   author = "Chaté, Hugues",
   title = "Dry Aligning Dilute Active Matter", 
   journal= "Annual Review of Condensed Matter Physics",
   year = "2020",
   volume = "11",
   number = "Volume 11, 2020",
   pages = "189-212",
   doi = "https://doi.org/10.1146/annurev-conmatphys-031119-050752",
   url = "https://www.annualreviews.org/content/journals/10.1146/annurev-conmatphys-031119-050752",
   publisher = "Annual Reviews",
   issn = "1947-5462",
   type = "Journal Article",
  }

@article{Kiely_2022,
   title={Superfluidity in the one-dimensional Bose-Hubbard model},
   volume={105},
   ISSN={2469-9969},
   url={http://dx.doi.org/10.1103/PhysRevB.105.134502},
   number={13},
   journal={Physical Review B},
   publisher={American Physical Society (APS)},
   author={Kiely, Thomas G. and Mueller, Erich J.},
   year={2022},
   month=apr }

@book{Giamarchi_2003,
    author = {Giamarchi, Thierry},
    title = {Quantum Physics in One Dimension},
    publisher = {Oxford University Press},
    year = {2003},
    month = {12},
    isbn = {9780198525004},
    url = {https://doi.org/10.1093/acprof:oso/9780198525004.001.0001},
}

@article{Penrose_1956,
  title = {Bose-Einstein Condensation and Liquid Helium},
  author = {Penrose, Oliver and Onsager, Lars},
  journal = {Phys. Rev.},
  volume = {104},
  issue = {3},
  pages = {576--584},
  numpages = {0},
  year = {1956},
  month = {Nov},
  publisher = {American Physical Society},
  url = {https://link.aps.org/doi/10.1103/PhysRev.104.576}
}

@article{Haldane_1981,
  title = {Effective Harmonic-Fluid Approach to Low-Energy Properties of One-Dimensional Quantum Fluids},
  author = {Haldane, F. D. M.},
  journal = {Phys. Rev. Lett.},
  volume = {47},
  issue = {25},
  pages = {1840--1843},
  numpages = {0},
  year = {1981},
  month = {Dec},
  publisher = {American Physical Society},
  doi = {10.1103/PhysRevLett.47.1840},
  url = {https://link.aps.org/doi/10.1103/PhysRevLett.47.1840}
}

@article{Antonov_2025,
   title={Engineering active motion in quantum matter},
   volume={7},
   ISSN={2643-1564},
   url={http://dx.doi.org/10.1103/z3gm-32jn},
   number={3},
   journal={Physical Review Research},
   publisher={American Physical Society (APS)},
   author={Antonov, Alexander P. and Zheng, Yuanjian and Liebchen, Benno and Löwen, Hartmut},
   year={2025},
   month=jul }

@article{Nadonly_2025,
  title = {Nonreciprocal Synchronization of Active Quantum Spins},
  author = {Nadolny, Tobias and Bruder, Christoph and Brunelli, Matteo},
  journal = {Phys. Rev. X},
  volume = {15},
  issue = {1},
  pages = {011010},
  numpages = {21},
  year = {2025},
  month = {Jan},
  publisher = {American Physical Society},
  doi = {10.1103/PhysRevX.15.011010},
  url = {https://link.aps.org/doi/10.1103/PhysRevX.15.011010}
}

@article{penner_2025,
  title = {Heat-to-motion conversion for quantum active matter},
  author = {Penner, Alexander-Georg and Viotti, Ludmila and Fazio, Rosario and Arrachea, Liliana and von Oppen, Felix},
  journal = {Phys. Rev. B},
  volume = {112},
  issue = {18},
  pages = {L180303},
  numpages = {6},
  year = {2025},
  month = {Nov},
  publisher = {American Physical Society},
  doi = {10.1103/r6tm-nx19},
  url = {https://link.aps.org/doi/10.1103/r6tm-nx19}
}

@article{adachi_2022,
  title = {Activity-induced phase transition in a quantum many-body system},
  author = {Adachi, Kyosuke and Takasan, Kazuaki and Kawaguchi, Kyogo},
  journal = {Phys. Rev. Res.},
  volume = {4},
  issue = {1},
  pages = {013194},
  numpages = {23},
  year = {2022},
  month = {Mar},
  publisher = {American Physical Society},
  doi = {10.1103/PhysRevResearch.4.013194},
  url = {https://link.aps.org/doi/10.1103/PhysRevResearch.4.013194}
}

@article{Taskan_2024,
  title = {Activity-induced ferromagnetism in one-dimensional quantum many-body systems},
  author = {Takasan, Kazuaki and Adachi, Kyosuke and Kawaguchi, Kyogo},
  journal = {Phys. Rev. Res.},
  volume = {6},
  issue = {2},
  pages = {023096},
  numpages = {14},
  year = {2024},
  month = {Apr},
  publisher = {American Physical Society},
  doi = {10.1103/PhysRevResearch.6.023096},
  url = {https://link.aps.org/doi/10.1103/PhysRevResearch.6.023096}
}

@misc{yuan_2024,
      title={Quantum Analog of Vicsek Model for Active Matter}, 
      author={Hong Yuan and L. X. Cui and L. T. Chen and C. P. Sun},
      eprint={2407.09860},
      archivePrefix={arXiv},
            year={2026}
}

@Article{Bricard_2013,
author={Bricard, Antoine
and Caussin, Jean-Baptiste
and Desreumaux, Nicolas
and Dauchot, Olivier
and Bartolo, Denis},
title={Emergence of macroscopic directed motion in populations of motile colloids},
journal={Nature},
year={2013},
month={Nov},
day={01},
volume={503},
number={7474},
pages={95-98},
issn={1476-4687},
doi={10.1038/nature12673},
url={https://doi.org/10.1038/nature12673}
}

@Article{Geyer_2018,
author={Geyer, Delphine
and Morin, Alexandre
and Bartolo, Denis},
title={Sounds and hydrodynamics of polar active fluids},
journal={Nature Materials},
year={2018},
month={Sep},
day={01},
volume={17},
number={9},
pages={789-793},
issn={1476-4660},
doi={10.1038/s41563-018-0123-4},
url={https://doi.org/10.1038/s41563-018-0123-4}
}

@article{Kaiser_2017,
author = {Andreas Kaiser  and Alexey Snezhko  and Igor S. Aranson },
title = {Flocking ferromagnetic colloids},
journal = {Science Advances},
volume = {3},
number = {2},
pages = {e1601469},
year = {2017},
doi = {10.1126/sciadv.1601469},
URL = {https://www.science.org/doi/abs/10.1126/sciadv.1601469},
}

@Article{Shaebani_2020,
author={Shaebani, M. Reza
and Wysocki, Adam
and Winkler, Roland G.
and Gompper, Gerhard
and Rieger, Heiko},
title={Computational models for active matter},
journal={Nature Reviews Physics},
year={2020},
month={Apr},
day={01},
volume={2},
number={4},
pages={181-199},
issn={2522-5820},
doi={10.1038/s42254-020-0152-1},
url={https://doi.org/10.1038/s42254-020-0152-1}
}

@BOOK{Gardiner_2004,
  title     = "Quantum Noise : a Handbook of Markovian and {Non-Markovian}
               Quantum Stochastic Methods with Applications to Quantum Optics",
  author    = "Gardiner, C W and Zoller, P",
  publisher = "Springer",
  year      =  2004,
}

@article{khasseh_2024,
  title = {Active Quantum Flocks},
  author = {Khasseh, Reyhaneh and Wald, Sascha and Moessner, Roderich and Weber, Christoph A. and Heyl, Markus},
  journal = {Phys. Rev. Lett.},
  volume = {135},
  issue = {24},
  pages = {248302},
  numpages = {8},
  year = {2025},
  month = {Dec},
  publisher = {American Physical Society},
  doi = {10.1103/rd46-hr3q},
  url = {https://link.aps.org/doi/10.1103/rd46-hr3q}
}

@misc{steiner2026,
      title={Active quantum matter from monitored pure-state dynamics}, 
      author={Jacob F. Steiner and Felix von Oppen and Reinhold Egger},
      year={2026},
      eprint={2603.12629},
      archivePrefix={arXiv},
      primaryClass={quant-ph},
      url={https://arxiv.org/abs/2603.12629}, 
}

@misc{burgardt2026,
      title={Quantum-enabled active matter at the atomic scale}, 
      author={Sabrina Burgardt and Julian Feß and Alexander Guthmann and Silvia Hiebel and Aritra K. Mukhopadhyay and Sangyun Lee and Michael te Vrugt and Benno Liebchen and Hartmut Löwen and Raphael Wittkowski and Artur Widera},
      year={2026},
      eprint={2606.24615},
      archivePrefix={arXiv},
      primaryClass={quant-ph},
      url={https://arxiv.org/abs/2606.24615}, 
}

@article{Antonov_2026,
  title = {Modeling dissipation in quantum active matter},
  author = {Antonov, Alexander P. and Lee, Sangyun and Liebchen, Benno and L\"owen, Hartmut and Melles, Jannis and Morigi, Giovanna and Tuchkov, Yehor and te Vrugt, Michael},
  journal = {Phys. Rev. A},
  volume = {114},
  issue = {1},
  pages = {012201},
  numpages = {14},
  year = {2026},
  month = {Jul},
  publisher = {American Physical Society},
  doi = {10.1103/4gyc-3pht},
  url = {https://link.aps.org/doi/10.1103/4gyc-3pht}
}

@misc{brunner2026,
      title={Quantum motility-induced phase separation}, 
      author={Laurin Brunner and Ricard Alert and Reyhaneh Khasseh and Markus Heyl},
      year={2026},
      eprint={2608.25649},
      archivePrefix={arXiv},
      primaryClass={quant-ph},
      url={https://arxiv.org/abs/2608.25649}, 
}

@misc{mykey,
  note = {See the Supplemental Material for details of the derivation, the homogeneous approximation, and the analytical calculations of the phase boundary and eigenvectors.}
}

\begin{widetext}
\section{End Matter \label{sec:endmatter}}
\end{widetext}

\paragraph{Derivation of hydrodynamic equation of correlation functions---}
\noindent Following is a step-by-step guidelines for the derivation of hydrodynamics of two-point correlation functions (detailed derivation presented in \cite{mykey}). Starting from the Lindblad master equation, one can compute the time evolution of the expectation value of an arbitrary operator $\hat{\mathcal{O}}$ as follows:
\begin{equation}
    \begin{split}
\frac{d  }{dt}\big \langle \hat{\mathcal{O}} \big \rangle&=-i \big \langle \big [H,\hat{\mathcal{O}} \big ] \big \rangle + \text{Tr}\left \{ \hat{\mathcal{O}} \big ( \mathcal{D}_{\mathcal{M}}(\hat{\rho}) +\mathcal{D}_{\mathcal{A}}(\hat{\rho}) \big )\right \} .
\label{eq:lindblad2}
    \end{split}
\end{equation}
Applying this to $ \big \langle \hat{\Pi}_{ij}^{\sigma_{1} \sigma_{2} }   \big \rangle $, we identify three distinct contributions: (1) the coherent evolution due to the Hamiltonian $ \hat{\mathcal{C}}_{H}^{\sigma_{1} \sigma_{2} } $, (2) dissipative dynamics due to active motion $\hat{\mathcal{C}}_{\mathcal{M}} ^{\sigma_{1},\sigma_{2}} $, and (3) dissipative dynamics due to the alignment interactions $\hat{\mathcal{C}}_{\mathcal{A}} ^{\sigma_{1},\sigma_{2}} $. 
%
%

Using the commutation/anti-commutation relations for hardcore bosons,
$
\big [  \hat{c}_{j,\sigma_{1}},\hat{c}_{i,\sigma_{2}}^{\dagger} \big ]=(1-2\hat{n}_{i,\sigma_1})\delta_{i,j} \delta_{\sigma_{1},\sigma_{2}}$
and 
$
\big \{  \hat{c}_{j,\sigma_1},\hat{c}_{i,\sigma_2}^{\dagger} \big \}=\delta_{\sigma_{1},\sigma_{2}}$ for $ i = j$, and
 the cyclic property of trace, we derive closed equations of motion for $ \langle \hat{\Pi}_{ij}^{\sigma_{1} \sigma_{2}}  \rangle$:
%
%
\begin{equation}
    \begin{split}
\frac{d}{dt} \big \langle \hat{\Pi}_{ij}^{\sigma_{1} \sigma_{2} } \big \rangle
=&
ig
\big \langle    \hat{\mathcal{C}}_{H}^{\sigma_{1} \sigma_{2} }  \big \rangle 
+
\frac{\Gamma_{\mathcal{M}}}{2}
\big \langle    \hat{\mathcal{C}}_{\mathcal{M}} ^{\sigma_{1} \sigma_{2} } 
 \big \rangle 
+
\frac{\Gamma_{\mathcal{A}}}{2}
\big \langle    \hat{\mathcal{C}}_{\mathcal{A}} ^{\sigma_{1} \sigma_{2} }  
 \big \rangle ,
\label{eq:main1}
    \end{split}
\end{equation}
where $\hat{\mathcal{C}}_{H,\mathcal{M},\mathcal{A}} $ denote the respective contributions as mentioned before. 


The coherent Hamiltonian evolution of the system arises from local spin-flip dynamics, which contribute to the dynamics of  
correlation functions through the term $\hat{\mathcal{C}}_{H}^{\sigma_{1} \sigma_{2} } $, given by
\begin{equation}
    \begin{split}
 \hat{\mathcal{C}}_{H}^{\sigma_{1} \sigma_{2} } =& 
 \hat{\Pi}_{i,j}^{\sigma_{1} \bar{\sigma}_{2}} (1-2\hat{n}_{j,\sigma _{2}})   -  \hat{\Pi}_{i,j}^{\bar{\sigma}_{1}\sigma_{2}} (1-2\hat{n}_{i,\sigma_{1}})  .
\label{eq:c_h}
    \end{split}
\end{equation}
The quantum nature of the flocking state is directly tied to the spin-flip amplitude $g$. In the absence of quantum superposition ($g=0$), the system evolves solely through dissipative processes. In this limit, no quantum superpositions are generated and the long-distance coherence vanishes,  i.e., $C(t)=0$.

The explicit form of $ \hat{\mathcal{C}}_{\mathcal{M}}  ^{\sigma_{1} \sigma_{2} }$, which is the contribution due to the dissipative active motion, is given as 
%
\begin{equation}
    \begin{split}
 \hat{\mathcal{C}}_{\mathcal{M}}  ^{\sigma_{1} \sigma_{2} }
 = &
 \Big [ \sum _{s= \pm 1 } \left ( s \hat{n}_{j-s \sigma_{2},\sigma_{2} } +   s \hat{n}_{i-s \sigma_{1} ,\sigma_{1} } \right )-2 \Big ]
 \hat{\Pi}_{ij}^{\sigma_{1} \sigma_{2}} 
\label{eq:c_m}
    \end{split}
\end{equation}
where $\hat{n}_{i,\sigma}$ is the number operator at lattice site $i$ and spin $\sigma$.
The contribution from dissipative alignment interactions, $\hat{\mathcal{C}}_{A} ^{\sigma_{1} \sigma_{2} }$, can be expressed as
\begin{equation}
    \begin{split}
    \scriptstyle
\hat{\mathcal{C}}_{\mathcal{A}} ^{\sigma _{1}\sigma_{2}}
  =&
\Big [\frac{\kappa}{r} \big ( \hat{F}_{j}^{\sigma_{2}}(1/4) + \hat{F}_{i}^{\sigma_{1}}(-1/4)     + \sigma _{2} \hat{M}_{j} + \sigma _{1} \hat{M}_{i} -
   \sigma _{1} \sigma_{2}\big )  
\\
 & 
  + \left ( \frac{\kappa}{2r} \right )^2
  \big (\hat{G}_{j}^{\sigma_{1} \sigma_{2}}(1) + \hat{G}_{i}^{\sigma_{2} \sigma_{1}}(-1)  
 -\hat{M}_{i} ^{2}  
  -
 \hat{M}_{j}  ^{2}  
  \\
  & 
  +
 2\sigma_{1}\hat{M}_{j}  \big )
 -2
 \Big ]
 \hat{\Pi}_{ij}^{\sigma_{1} \sigma_{2}} ,
\label{eq:c_a}
    \end{split}
\end{equation}
where the auxiliary functions $\hat{F}$ and $\hat{G}$ are defined as $\hat{F}^{\sigma }_{a}(\alpha)=(\hat{\rho}_{a}+\bar{\sigma} \hat{m}_{a}) (-\sigma\hat{M}_{a}+\alpha)$ and $\hat{G}^{\sigma_{1} \sigma_{2} }_{a}(\alpha)=(\hat{\rho}_{a}+ \bar{\sigma}_{2} \hat{m}_{a}) (-\sigma_{1}\hat{M}_{a}+\alpha)$, respectively, and $a \in \{i,j\}$. The effect of the local environment enters through $\hat{M}_{i}=\sum _{|j|=1}^{r}\hat{m}_{j+i}$, which measures surrounding magnetization at the lattice site $i$ within the interaction radius $r$.

 \paragraph{Homogeneous approximation to the dynamics---}
\noindent We assume spatial homogeneity as $\left \langle \hat{\rho}_{i}  \right \rangle =\rho(t) $, $\left \langle \hat{m}_{i} \right \rangle =m(t) $, $ \langle \hat{f}_{i}\rangle =f(t) $, and $\left \langle M_{i} \right \rangle =\bar{M} $. 
Homogeneous treatment of each contribution can be determined, and $\big \langle \hat{\mathcal{C}}_{H} ^{\sigma_1 \sigma_2}\big \rangle $ is given by,
\begin{equation}
    \begin{split}
\big \langle \hat{\mathcal{C}}_{H} ^{\sigma_1 \sigma_2}\big \rangle 
=&
-\left ( 1+ 2 \left \langle  \hat{n}_{j,\sigma _{2}} \right \rangle \right )
\big \langle    \hat{\Pi}_{ij}^{\sigma_{1}  \bar{\sigma_{2}}}   \big \rangle 
\\
&
+
\left ( 1 +2 \left \langle  \hat{n}_{i,\sigma _{1}} \right \rangle \right )
\big \langle    \hat{\Pi}_{ij}^{ \bar{\sigma_{1}} \sigma_{2} }   \big \rangle 
\\
&
+2
\left ( \big \langle    \hat{\Pi}_{jj}^{\sigma_{2}  \bar{\sigma_{2}}}  \big \rangle  - \big \langle     \hat{\Pi}_{ii}^{ \bar{\sigma_{1}} \sigma_{1} }  \big \rangle \right ) \big \langle     \hat{\Pi}_{ij}^{ \sigma_{1} \sigma_{2} }  \big \rangle   .
\label{eq:c_eq5}
    \end{split}
\end{equation}
To get the last term in Eq. \eqref{eq:c_eq5}, we assume homogeneity as $\big \langle    \hat{\Pi}_{jj}^{\sigma_{2}  \bar{\sigma_{2}}}  \big \rangle =  \Pi ^{\sigma_{2}  \bar{\sigma_{2}}} $, which is independent of lattice sites. This gives 
\begin{equation}
    \begin{split}
 \big \langle    \hat{\Pi}_{jj}^{\sigma_{2}  \bar{\sigma_{2}}}  \big \rangle  - \big \langle     \hat{\Pi}_{ii}^{ \bar{\sigma_{1}} \sigma_{1} }  \big \rangle = \begin{cases}
  -if \quad \text{when } \quad \sigma_{1,2} =+\\      
   if \quad \text{when } \quad \sigma_{1,2} =-\\
  0 \quad \text{when } \quad \sigma_1 \ne \sigma_2 ,
\end{cases}
    \end{split}
\end{equation}
where $f$ is the homogeneous spin-flip current.

Doing the same factorization for $\big \langle \hat{\mathcal{C}}_{\mathcal{M}}^{\sigma_{1} \sigma_{2} }  \big \rangle $, gives
\begin{equation}
    \begin{split}
\big \langle \hat{\mathcal{C}}_{\mathcal{M}}^{\sigma_{1} \sigma_{2}}  \big \rangle  
= &
-2
  \big \langle \hat{\Pi}_{ij}^{\sigma_{1} \sigma_{2} }  \big \rangle  .
    \end{split}
\end{equation}
The correlation functions depend on the distance between the lattice sites, and in this light, all other terms can be canceled.

Similarly, we do the homogeneous approximation of $\big \langle \hat{\mathcal{C}}_{\mathcal{A}} ^{\sigma_{1} \sigma_{2}}  \big \rangle $: first
we split $\hat{\mathcal{C}}_{\mathcal{A}} ^{\sigma_{1} \sigma_{2}}$ as 
\begin{equation}
    \begin{split}
 \hat{\mathcal{C}}_{\mathcal{A}} ^{\sigma_{1} \sigma_{2}} 
  =&
  -2   \big \langle \hat{\Pi}_{ij}^{\sigma_{1} \sigma_{2} }  \big \rangle 
  +
  \frac{\kappa}{r}
 \hat{\mathcal{C}}_{\mathcal{A}(1)} ^{\sigma_{1} \sigma_{2}} 
  +
  \left (  \frac{\kappa}{r} \right )^2 
\hat{\mathcal{C}}_{\mathcal{A}(2)} ^{\sigma_{1} \sigma_{2}} ,
    \end{split}
\end{equation}
where $\hat{\mathcal{C}}_{\mathcal{A}(1)} ^{\sigma_{1} \sigma_{2}}$ and $\hat{\mathcal{C}}_{\mathcal{A}(2)} ^{\sigma_{1} \sigma_{2}}$ are the linear and quadratic order coefficients, respectively.
After replacing expectation with homogeneous quantities, we get
\begin{widetext}
\begin{equation}
    \begin{split}
\big \langle  \hat{\mathcal{C}}_{\mathcal{A}(1)} ^{\sigma _{1}\sigma_{2}}  \big \rangle 
     =&    
  \big ( -\sigma _{1} \sigma_{2}  +  \sigma _{2}   \bar{M} +  \sigma_{1}  \bar{M}  \big )  \left \langle \hat{\Pi}_{ij}^{\sigma_1 \sigma_2} \right \rangle 
+ 
       \left (   \big (- \sigma _{1} \bar{M}   + \frac{1}{4}\big )\left ( \rho  +\bar{\sigma_{2}} m   \right)
   +
   \big ( -\sigma_{2}  \bar{M}  -\frac{1}{4}\big ) \left (\rho + \bar{\sigma_{1}} m \right )\right ) 
   \left \langle \hat{\Pi}_{ij}^{\sigma_1 \sigma_2} \right \rangle 
   \\
     &
     +
      \big (- \sigma _{2}  \bar{M}  + \frac{1}{4}\big )   \left \langle \hat{\Pi}_{jj}^{\bar{\sigma_2} \sigma_2} \right \rangle    \left \langle \hat{\Pi}_{ij}^{\sigma_1 \bar{\sigma_2}} \right \rangle 
   +
   \big ( -\sigma_{1}  \bar{M}   -\frac{1}{4}\big )   \left \langle \hat{\Pi}_{ij}^{\bar{\sigma_1} \sigma_2} \right \rangle     \left \langle \hat{\Pi}_{ii}^{\sigma_1 \bar{\sigma_1}} \right \rangle ,
   \\
    \end{split}
\end{equation}
\begin{equation}
    \begin{split}
\left \langle \hat{\mathcal{C}}_{\mathcal{A}(2)} ^{\sigma_{1} \sigma_{2}}  \right \rangle
=&
  \big ( 
 - 2\bar{M} ^{2}
  +
2\sigma_{1} \bar{M}    \big ) \left \langle  \hat{\Pi}_{ij}^{\sigma_{1} \sigma_{2} } \right \rangle 
 +
   \big (
\left ( -\sigma_{1}  \bar{M}  +1\right )  \left (\rho  + \bar{\sigma _{2}} m\right ) 
  + 
  \left (-\sigma_{2} \bar{M}  -1 \right) \left (\rho + \bar{\sigma_{1}} m\right ) \big )  \left \langle \hat{\Pi}_{ij}^{\sigma_{1} \sigma_{2} } 
  \right \rangle
  \\
  &
  +  
\left ( -\sigma_{1} \bar{M}   +1\right )  \left \langle \hat{\Pi}_{jj}^{\bar{\sigma_2} \sigma_2} \right \rangle    \left \langle \hat{\Pi}_{ij}^{\sigma_1 \bar{\sigma_2}} \right \rangle 
  + 
  \left (-\sigma_{2}  \bar{M}  -1 \right)  \left \langle \hat{\Pi}_{ij}^{\bar{\sigma_1} \sigma_2} \right \rangle     \left \langle \hat{\Pi}_{ii}^{\sigma_1 \bar{\sigma_1}} \right \rangle   .
    \end{split}
\end{equation}
\end{widetext}

Finally, the dynamics of the correlation functions with the homogeneous approximation can be written as 
\begin{equation}
    \begin{split}
    \frac{d}{dt}\mathbf{\Pi}
    =
    \mathcal{K} \cdot \mathbf{\Pi}
    +
    \mathcal{O}(m^{3}),
    \label{eq:rho_mat1}
    \end{split}
\end{equation}
where matrix $\mathcal{K}$ is given as
\begin{equation}
\mathcal{K}
=
\begin{pmatrix}
\phantom{\hat{I}_{.}} c_{+} &  0 & s_{1+} & s_{1-} \\
\phantom{\hat{I}_{.}} 0 &  c_{-} & s_{2+} & s_{2-} \\
\phantom{\hat{I}_{.}} s_{3+} & s_{3-} & d_{+} & 0 \\
\phantom{\hat{I}_{.}} s_{4+} & s_{4-} & 0 & d_{-}
\end{pmatrix}
.
\end{equation}
We considered all contributions up to second order in the magnetization $m$. Accordingly, the coefficients entering the dynamical matrix $\mathcal{K}$ are functions of the homogeneous magnetization $m$, the density $\rho$ (which we fix to average value $\rho = 1/2$), the spin-flip current $f$, and the local energy $\epsilon$. Neglecting contributions that are sub-leading in the interaction radius $r$, the relevant constants take the form
\begin{equation}
\begin{split}
c_{\pm }= & \mp 
2gf- \Gamma_{\mathcal{M}} - \Gamma_{\mathcal{A}} \left [ 1 + \frac{\kappa}{2r} \mp \kappa m 
 - m^2  \big ( \kappa^{2} +2\kappa  \big )  \right ]  ,
 \\
     d_{\pm}= &
- \Gamma_{\mathcal{M}} 
- \Gamma_{\mathcal{A}} \left [ 1 -\frac{\kappa}{2r} +m^2 \big ( \kappa^2  \pm \kappa \big )\right ] ,
\end{split}
\end{equation}
\begin{equation}
\begin{split}
      s_{1\pm }=&
\frac{1}{2} \left [ \pm ig \big  (1 - 2 m \big ) 
-
\Gamma_{\mathcal{A}}\kappa m \big ( \epsilon \pm i f \big ) \right ] ,
\\
    s_{2 \pm}
=&
\frac{1}{2}\left [
\mp ig \big  (1 + 2m \big ) 
+
\Gamma_{\mathcal{A}}\kappa m\big ( \epsilon \pm i f\big )
\right ] ,
\\
    s_{3 \pm }= &
\frac{1}{2}\left [
ig \big  ( \pm 1- 2m \big ) 
\pm 
\Gamma_{\mathcal{A}}\kappa m\big ( \epsilon +if \big )
\right  ] ,
\\
    s_{4 \pm }= &
\frac{1}{2}\left [
ig \big  ( \mp 1 - 2m \big )
\pm 
 \Gamma_{\mathcal{A}}\kappa m\big ( \epsilon -if\big )
\right ],
\end{split}
\end{equation}
where we have also used $\bar{M}/2r =m$, by setting maximum possible value for interaction radius $r=L/2$.
We get the Eq. \eqref{eq:matrix_form} by setting $\Gamma_{\mathcal{M}/\mathcal{A}}=\Gamma$ and neglecting $1/r$ contributions. Without loss of generality, we set the steady-state value of $\epsilon$ to zero, as it's dynamics decouples from the magnetization and spin-flip current dynamics \cite{mykey}.

\begin{figure}
    \centering
 \includegraphics[clip,trim=0cm 0cm 0cm 0cm,width=.3\textwidth]{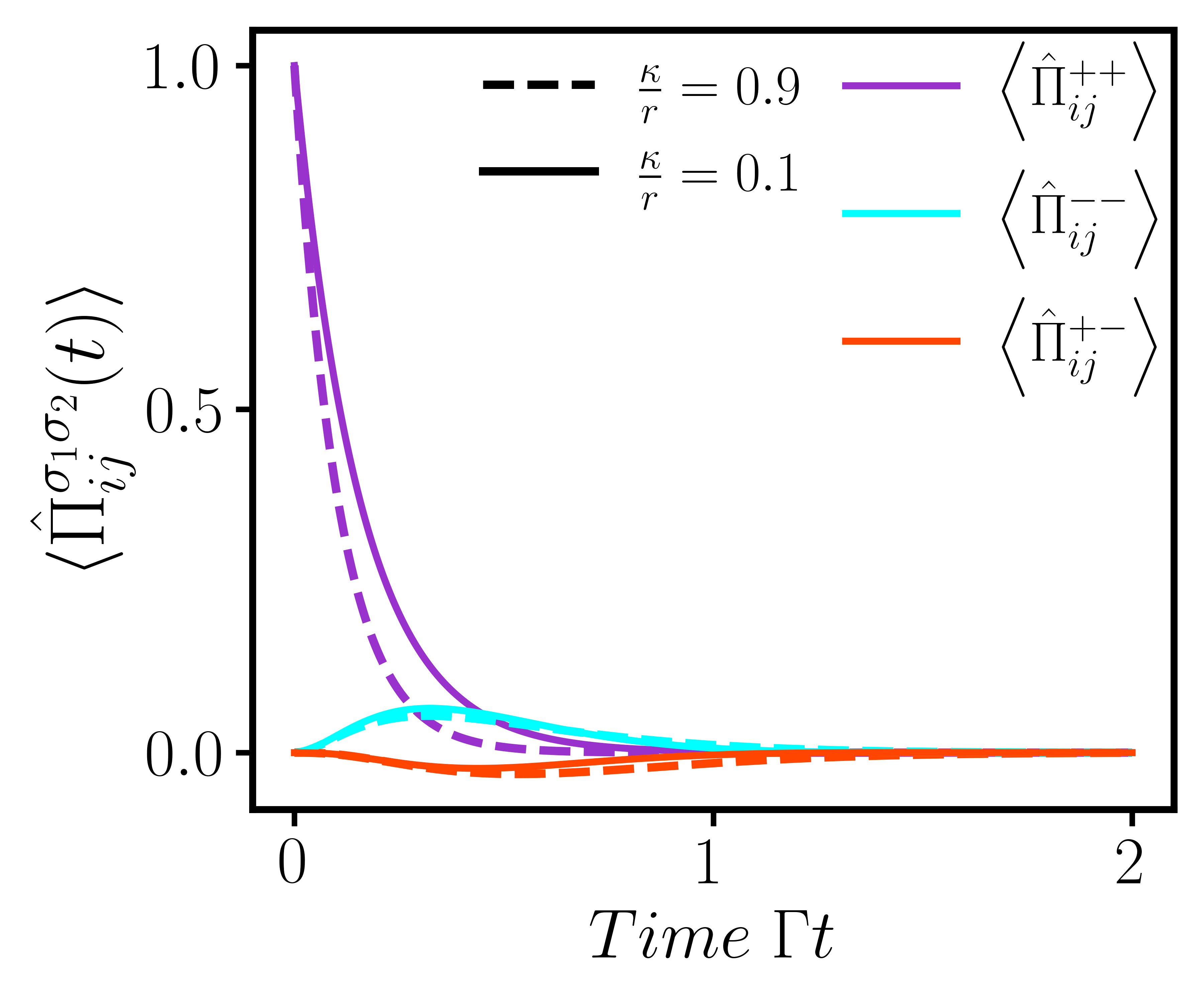}
\caption{Dynamics of correlation functions in disordered phase from Eq. \eqref{eq:rho_mat_dis}, and parameters are same as FIG. \ref{fig:dynamics}.} 
    \label{fig:dynamics2}
\end{figure}

 \paragraph{Decaying solution in the disordered state---}
\noindent
In the disordered state, the dynamics of the correlation function is governed by a matrix $\mathcal{K}_0=\mathcal{K}(m=0)$ as
\begin{equation}
    \begin{split}
    \frac{d}{dt}\mathbf{\Pi}
    =
    \mathcal{K}_{0} \cdot \mathbf{\Pi},
    \quad 
    \mathcal{K}_{0}
    =
    \begin{pmatrix}
\phantom{\hat{I}_{.}} c_{+}& 0& s& -s\\
\phantom{\hat{I}_{.}} 0& c_{+}& -s& s\\
\phantom{\hat{I}_{.}} s& -s& c_{-}& 0\\
\phantom{\hat{I}_{.}} -s& s& 0& c_{-}
\end{pmatrix}
,
     \label{eq:rho_mat_dis}
 \end{split}
\end{equation}
where the constants are given by
\begin{equation}
\begin{split}
    c_{\pm}=& -
\Gamma_{\mathcal{M}} - \Gamma_{\mathcal{A}} \big [ 1 \pm  \frac{\kappa}{2r}   \big ]  ,
\\
      s=&
\frac{ig}{2}  .
\end{split}
\end{equation}
The solutions are shown in FIG.~\ref{fig:dynamics2}(top) for various values of $\kappa/r$, where the correlations rapidly decay to zero. 

These solutions can diverge when $\kappa/r \gg 1$, but in this regime the homogeneous approximation also becomes invalid. In other words, there exists a lower bound on the interaction radius required to sustain the flocking state \cite{khasseh_2024}. This implies that for a system exhibiting a flocking steady state, $\kappa$ must be small or comparable to $r$, in which case the correlation functions in the disordered phase always decay to zero.
Assuming a large interaction radius ($r \sim L/2$) within the mean-field approximation, one can approximate $1 + \frac{\kappa}{2r} \sim 1$, allowing an exact solution for the correlation dynamics:
\begin{equation}
    \begin{split}
        \mathbf{\Pi}(t)=
        \frac{1}{2}e^{-\left (\Gamma_{\mathcal{M}}+\Gamma_{\mathcal{A}}\right ) t}
        \begin{pmatrix}
            \cos gt + 1 \\
            \cos g  t - 1 \\
            i \sin g  t \\
           i \sin g t
        \end{pmatrix}
    \end{split}
\end{equation}
This solution is obtained by taking a particular initial condition where $\big \langle \hat{\Pi}_{ij}^{++} (t=0)\big \rangle=1$ with all other correlations set to zero at $t=0$.

\clearpage                
\onecolumngrid            
\clearpage                



\section*{Supplementary material}

# Supplementary material for "Superfluidity in active quantum flocks"

Byjesh N. Radhakrishnan, Reyhaneh Khasseh, Thomas L. Schmidt, and Markus Heyl

**Contents**



This supplementary material presents the main calculations in detail. To make this self-consistent, we start with the coarse-grained dynamics of magnetization and density in the Section I. In Section II, we derive the full hydrodynamic expression for the one-body correlation function. Section III is devoted to applying homogeneous approximations to obtain the final form of the hydrodynamic equations. Finally, in Section IV, we determine the condition under which the largest eigenvalue ($\lambda_{max}$) of the matrix $\mathcal{K} = \mathcal{K}_0 + \mathcal{K}_m$ vanishes; this result defines the phase boundary between correlated and uncorrelated motion. In the final Section, we details the numerical calculation of eigenvector of $\mathcal{K}$ matrix and derive the analytical expression for eigenvector.

## I. Coarse-grained dynamics of magnetization and spin-flip current.

We develop a coarse-grained equation of motion for the evolution of density, magnetization, local spin-flip current starting from the Lindblad master equation, which follows as [1]

$$\frac{d}{dt}\langle \hat{\rho}_l \rangle = \Gamma_{\mathcal{M}} \left[ \langle (\frac{1}{2} - \hat{\rho}_l)\delta \hat{m}_l \rangle - \langle \hat{m}_l \delta \hat{\rho}_l \rangle + \frac{1}{2} \langle \delta^2 \hat{\rho}_l \rangle \right], \tag{1}$$

$$\frac{d}{dt}\langle \hat{m}_l \rangle = g\langle \hat{f}_l \rangle - 2\Gamma_{\mathcal{A}} \langle \hat{P}_l^\dagger \hat{P}_l \hat{m}_l \rangle - \Gamma_{\mathcal{M}} \left[ \langle (\frac{1}{2} - \hat{\rho}_l)\delta \hat{\rho}_l \rangle + \langle \hat{m}_l \delta \hat{m}_l \rangle + \frac{1}{2} \langle \delta^2 \hat{m}_l \rangle \right], \tag{2}$$

$$\frac{d}{dt}\langle \hat{f}_l \rangle = -4g \langle \hat{m}_l \rangle - \frac{\Gamma_{\mathcal{A}}}{2} \langle \{ \hat{P}_l^\dagger \hat{P}_l, \hat{f}_l \} \rangle - \Gamma_{\mathcal{M}} \left[ \langle (1 + \delta \hat{m}_l) \hat{f}_l \rangle \right], \tag{3}$$

where we used the finite difference formula for first and second order derivatives, defined as

$$\delta \hat{\mathcal{O}} = \hat{\mathcal{O}}_{l+1} - \hat{\mathcal{O}}_{l-1}, \quad \text{and} \quad \delta^2 \hat{\mathcal{O}} = \hat{\mathcal{O}}_{l+1} + \hat{\mathcal{O}}_{l-1} - 2\hat{\mathcal{O}}_l. \tag{4}$$

We analyze the solutions to the coarse-grained dynamics [Eqs. (1)-(3)] to see the different phase behaviors. Within the homogeneous phase approximation, Eq. (1)–(3) reduces to

$$\frac{d}{dt} m(t) = g f(t) - \Gamma_{\mathcal{A}} \langle \hat{P}_l^\dagger \hat{P}_l \hat{m}_l \rangle, \tag{5}$$

$$\frac{d}{dt} f(t) = -4g m(t) - \frac{\Gamma_{\mathcal{A}}}{2} \langle \{ \hat{P}_l^\dagger \hat{P}_l, \hat{f}_l \} \rangle - \Gamma_{\mathcal{M}} f(t). \tag{6}$$

To solve dynamics in Eqs. (5) and (6) in the steady state, one has to deal with the non-linearity $\langle \hat{P}_l^\dagger \hat{P}_l \hat{m}_l \rangle$. We expand $\hat{P}_l$ term in orders of $\hat{m}_l$ (keeping in mind that $\hat{P}_l$ has a general expression $\hat{P}_l = \exp(-\frac{\kappa}{2r} \hat{m}_l \sum_{|j|=1}^{r} \hat{m}_{l+j})$ [1]) and keep terms upto the second order in $\hat{m}_l$:

$$\langle \hat{P}_l^\dagger \hat{P}_l \hat{m}_l \rangle = m - 2\kappa \langle \hat{m}_l^2 \hat{M}_l \rangle + 2\kappa^2 \langle \hat{m}_l^3 \hat{M}_l \hat{M}_l \rangle, \tag{7}$$

where we have used maximum possible interaction radius $r = L/2$, and defined the surrounding magnetization as $\hat{M}_l = \sum_{j \neq l} \hat{m}_j$. It's also worth noting here that all the local magnetization are independently fluctuating, and we

neglect all the cross-correlations. The fluctuations in the mean magnetization will be small for a large system size and can be neglected: hence, we can consider $\langle \hat{M}_l \rangle = m$. So, we get

$$\langle \hat{P}_l^\dagger \hat{P}_l \hat{m}_l \rangle = m - 2\kappa \langle \hat{m}_l^2 \rangle m + 2\kappa^2 \langle \hat{m}_l^3 \rangle m^2. \tag{8}$$

Now, fluctuations in the local magnetization should be considered: we follow the case studied in Active Ising models [2] and consider these fluctuations follow Gaussian statistics, which enable us to write the second and third cumulants of magnetization as follows:

$$\begin{aligned} \langle \hat{m}_l^2 \rangle &= m^2 + \sigma_m^2, \\ \langle \hat{m}_l^3 \rangle &= m^3 + qm, \end{aligned} \tag{9}$$

where $q$ and $\sigma_m$ are constants that contain microscopic details of the system. Now, let's first examine the steady state solution for spin flip-current, which can be found by setting the time derivative part equal to zero:

$$f \approx -\frac{4gm}{\Gamma_{\mathcal{M}} + \Gamma_{\mathcal{A}}}, \tag{10}$$

where we have used $\{\hat{P}_l^\dagger \hat{P}_l, \hat{f}_l\} = 2\hat{f}_l$, which is correct for first order expansions of $\hat{P}_l$.

Returning to the dynamics of magnetization and using the expression from Eq. (10) gives the non-trivial solution as

$$m^2 \approx \frac{1}{2\kappa_c^2} \frac{\kappa/\kappa_c - 1 - \Delta_g}{q - 1/\kappa_c}, \tag{11}$$

where we also assumed $|m| << 1$, which enables us to neglect the $m^4$ terms. The constant $\kappa_c = 1/2\sigma_m^2$ coming from the fluctuations of magnetization, and $\Delta_g = 4g^2/(\Gamma_{\mathcal{M}} + \Gamma_{\mathcal{A}})\Gamma_{\mathcal{A}}$ are the contribution from the spin-flip dynamics.

**Linear stability of flocking phase:** The solution of magnetization [Eq. (11)] is stable for either $\kappa > \kappa_c(1 + \Delta_g)$ and $q > 1/\kappa_c$ or $\kappa < \kappa_c(1 + \Delta_g)$ and $q < 1/\kappa_c$. In addition, linear stability analysis can make clear which one of these solutions is stable. For this purpose, we consider a small perturbation over a homogeneous state $m(t) = m_0 + \Delta m$, and substituting back to Eq. (5) and keeping terms linear in $\Delta m$ gives

$$\frac{d}{dt}\Delta m(t) = -4\Gamma_{\mathcal{A}} \left( \frac{\kappa}{\kappa_c} - 1 \right) \Delta m(t), \tag{12}$$

where we also set $g = 0$. The solution for this equation will be an exponential function, either growth or decay, depending on the sign of $\kappa/\kappa_c - 1$. This implies that the homogeneous state is stable only if $\kappa > \kappa_c$ (or $\kappa > \kappa_c(1 + \Delta_g)$ for $g \neq 0$).

**Phase diagram:** The phase diagram in the $\kappa$–$g$ plane, separating the disordered and flocking phases, is constructed by analyzing the modification of the magnetization induced by a finite field $g$. This introduces an additional contribution $\Delta_g$ to the magnetization, which in turn modifies the stability condition to $\kappa > \kappa_c(1 + \Delta_g)$. Consequently, the blue phase boundary in Fig. 2 is determined from this corrected stability criterion.

$$\kappa_c(g) = (1 + \Delta_g)\kappa_c(g = 0), \tag{13}$$

where $\kappa_c(g = 0)$ is set to 2.4, the value obtained by numerical fitting [1], in the plot.

## II. Derivation of time evolution of correlation function $\langle \hat{\Pi}_{lq}^{\sigma_1\sigma_2} \rangle$

Here, we derive the time evolution of $\left\langle \hat{\Pi}_{lq}^{\sigma_1\sigma_2} \right\rangle$, and the resulting dynamics has three contributions: (1) the coherent evolution due to the Hamiltonian $\hat{\mathcal{C}}_H^{\sigma_1\sigma_2}$, (2) dissipative dynamics due to active motion $\hat{\mathcal{C}}_{\mathcal{M}}^{\sigma_1,\sigma_2}$, and (3) dissipative dynamics due to alignment interactions $\hat{\mathcal{C}}_{\mathcal{A}}^{\sigma_1,\sigma_2}$. We will calculate each of these contributions as follows.

### A. Contribution due to the coherent Hamiltonian evolution

The contribution from the Hamiltonian dynamics is given by

$$\begin{aligned}\hat{\mathcal{C}}_H^{\sigma_1\sigma_2} =& \frac{1}{ig}\left(-i[\hat{H}, \hat{c}^\dagger_{l,\sigma_1}\hat{c}_{q,\sigma_2}]\right)\\ =&\hat{c}^\dagger_{l,\sigma_1}(1-2\hat{n}_{q,\sigma_2})\hat{c}_{q,\bar{\sigma}_2} - \hat{c}^\dagger_{l,\bar{\sigma}_1}(1-2\hat{n}_{l,\sigma_1})\hat{c}_{q,\sigma_2}\end{aligned} \tag{14}$$

where we have used the commutation/anti-commutation relations for creation and annihilation operator of hardcore bosons

$$\begin{aligned}\left[\hat{c}_{j,\sigma_1}, \hat{c}^\dagger_{i,\sigma_2}\right] =&(1-2\hat{n}_{i,\sigma_1})\delta_{i,j}\delta_{\sigma_1,\sigma_2},\\ \left\{\hat{c}_{j,\sigma}, \hat{c}^\dagger_{i,\sigma}\right\} =&\delta_{\sigma_1,\sigma_2}, \quad \text{when} \quad i=j\end{aligned} \tag{15}$$

and commutator identity $[AB,CD] = A[B,C]D + [A,C]BD + CA[B,D] + C[A,D]B$.

### B. Contribution due to the dissipative self-propulsion

Contributions from dissipative active motion $\hat{\mathcal{C}}_{\mathcal{M}}^{\sigma_1,\sigma_2}$, are given by

$$\hat{\mathcal{C}}_{\mathcal{M}}^{\sigma_1,\sigma_2} = \sum_{k=1}^{L}\sum_{\sigma=+,-}\left(\hat{\mathcal{M}}^\dagger_{k,\sigma}\left[\hat{\Pi}^{\sigma_1\sigma_2}_{lq}, \hat{\mathcal{M}}_{k,\sigma}\right] + \left[\hat{\mathcal{M}}^\dagger_{k,\sigma}, \hat{\Pi}^{\sigma_1\sigma_2}_{lq}\right]\hat{\mathcal{M}}_{k,\sigma}\right) \tag{16}$$

Performing the commutations for $\sigma = +$,

$$\begin{aligned}\sum_{k=1}^{L}\hat{\mathcal{M}}^\dagger_{k,+}\left[\hat{c}^\dagger_{l,+}\hat{c}_{q,+}, \hat{\mathcal{M}}_{k,+}\right] &= \left(-\hat{n}_{q+1} - (1-\hat{n}_{l-1}) + \delta_{l,q+1}\right)\hat{\Pi}^{++}_{lq},\\ \sum_{k=1}^{L}\left[\hat{\mathcal{M}}^\dagger_{k,+}, c^\dagger_{l,+}c_{q,+}\right]\hat{\mathcal{M}}_{k,+} &= \left(-\hat{n}_{l+1} - (1-\hat{n}_{q-1}) - \delta_{l,q-1}\right)\hat{\Pi}^{++}_{lq},\end{aligned} \tag{17}$$

and the contributions vanish for $\sigma = -$. Similar calculations lead to the final expression

$$\hat{\mathcal{C}}_{\mathcal{M}}^{\sigma_1\sigma_2} = \left(\hat{n}_{q-\sigma_2,\sigma_2} - \hat{n}_{q+\sigma_2,\sigma_2} + \hat{n}_{l-\sigma_1,\sigma_1} - \hat{n}_{l+\sigma_1,\sigma_1} - 2\right)\hat{\Pi}^{\sigma_1\sigma_2}_{lq}. \tag{18}$$

### C. Contribution due to the dissipative alignment interaction

This contribution is given by,

$$\hat{\mathcal{C}}_{\mathcal{A}}^{\sigma_1\sigma_2} = \sum_{l=1}^{L}\sum_{\sigma=+,-}\left(\hat{\mathcal{A}}^\dagger_{l,\sigma}\left[\hat{\Pi}^{\sigma_1\sigma_2}_{lq}, \hat{\mathcal{A}}_{l,\sigma}\right] + \left[\hat{\mathcal{A}}^\dagger_{l,\sigma}, \hat{\Pi}^{\sigma_1\sigma_2}_{lq}\right]\hat{\mathcal{A}}_{l,\sigma}\right). \tag{19}$$

Using the expressions $\hat{\mathcal{A}}_{l\sigma} = \hat{c}^\dagger_{l\sigma}\hat{c}_{l\bar{\sigma}}\hat{\mathcal{P}}_l$ along with $\hat{\mathcal{P}}_l = 1 - (\kappa/2r)\hat{m}_l\hat{M}_l$, gives

$$\begin{aligned}\hat{\mathcal{C}}_{\mathcal{A}}^{\sigma_1\sigma_2} = \sum_{k=1}^{L}\Bigg(&\hat{\mathcal{A}}^\dagger_{k,+}\left[\hat{\Pi}^{\sigma_1\sigma_2}_{lq}, \hat{c}^\dagger_{k,+}\hat{c}_{k,-}\right]\hat{\mathcal{P}}_k + \hat{\mathcal{P}}^\dagger_k\left[\hat{c}^\dagger_{k,-}\hat{c}_{k,+}, \hat{\Pi}^{\sigma_1\sigma_2}_{lq}\right]\hat{\mathcal{A}}_{k,+}\\ &+ \hat{\mathcal{A}}^\dagger_{k,-}\left[\hat{\Pi}^{\sigma_1\sigma_2}_{lq}, \hat{c}^\dagger_{k,-}\hat{c}_{k,+}\right]\hat{\mathcal{P}}_k + \hat{\mathcal{P}}^\dagger_k\left[\hat{c}^\dagger_{k,+}\hat{c}_{k,-}, \hat{\Pi}^{\sigma_1\sigma_2}_{lq}\right]\hat{\mathcal{A}}_{k,-}\\ &+ \hat{\mathcal{A}}^\dagger_{k,+}\hat{c}^\dagger_{k,+}\hat{c}_{k,-}\left[\hat{\Pi}^{\sigma_1\sigma_2}_{lq}, \hat{\mathcal{P}}_k\right] + \left[\hat{\mathcal{P}}^\dagger_k, \hat{\Pi}^{\sigma_1\sigma_2}_{lq}\right]\hat{c}^\dagger_{k,-}\hat{c}_{k,+}\hat{\mathcal{A}}_{k,+}\\ &+ \hat{\mathcal{A}}^\dagger_{k,-}\hat{c}^\dagger_{k,-}\hat{c}_{k,+}\left[\hat{\Pi}^{\sigma_1\sigma_2}_{lq}, \hat{\mathcal{P}}_k\right] + \left[\hat{\mathcal{P}}^\dagger_k, \hat{\Pi}^{\sigma_1\sigma_2}_{lq}\right]\hat{c}^\dagger_{k,+}\hat{c}_{k,-}\hat{\mathcal{A}}_{k,-}\Bigg).\end{aligned} \tag{20}$$

For the sake of simplicity, we first derive $\hat{\mathcal{C}}_{\mathcal{A}}^{++}$.

### 1. Derivation of $\hat{\mathcal{C}}_{\mathcal{A}}^{++}$

From Eq. (20), terms that contain commutation $\left[\hat{\Pi}_{lq}^{++}, \hat{c}_{k,+}^{\dagger}\hat{c}_{k,-}\right]$ can be collected together and simplified as $-\hat{\mathcal{P}}_q^{\dagger}\hat{\mathcal{P}}_q\hat{\Pi}_{lq}^{++} - \hat{\mathcal{P}}_l^{\dagger}\hat{\mathcal{P}}_l\hat{\Pi}_{lq}^{++} - \hat{\mathcal{P}}_q^{\dagger}\left[\hat{\Pi}_{lq}^{++}, \hat{\mathcal{P}}_q\right] - \hat{\mathcal{P}}_l^{\dagger}\left[\hat{\Pi}_{lq}^{++}, \hat{\mathcal{P}}_l\right]$. Substituting this result back to Eq. (20), and rearranging gives:

$$\begin{aligned}\hat{\mathcal{C}}_{\mathcal{A}}^{\sigma_1\sigma_2} =& -\hat{\mathcal{P}}_q^{\dagger}\hat{\mathcal{P}}_q\hat{\Pi}_{lq}^{++} - \hat{\mathcal{P}}_l^{\dagger}\hat{\mathcal{P}}_l\hat{\Pi}_{lq}^{++} - \hat{\mathcal{P}}_q^{\dagger}\left[\hat{\Pi}_{lq}^{++}, \hat{\mathcal{P}}_q\right] - \hat{\mathcal{P}}_l^{\dagger}\left[\hat{\Pi}_{lq}^{++}, \hat{\mathcal{P}}_l\right] \\ &+ \sum_{k=1}^{L}\hat{\mathcal{P}}_k^{\dagger}\left(\hat{n}_{k,-} + \hat{n}_{k,+} - 2\hat{n}_{k,-}\hat{n}_{k,+}\right)\left[\hat{\Pi}_{lq}^{++}, \hat{\mathcal{P}}_k\right] \\ &+ \sum_{k=1}^{L}\left[\hat{\mathcal{P}}_k^{\dagger}, \hat{\Pi}_{lq}^{++}\right]\hat{\mathcal{P}}_k\left(\hat{n}_{k,-} + \hat{n}_{k,+} - 2\hat{n}_{k,-}\hat{n}_{k,+}\right).\end{aligned} \tag{21}$$

To perform further analysis, we first evaluate the commutation $\left[\hat{\Pi}_{lq}^{++}, \hat{\mathcal{P}}_k\right]$ as

$$\left[\hat{\Pi}_{lq}^{++}, \hat{\mathcal{P}}_k\right] = \frac{\kappa}{2r}\left(\sum_{|j|=1}^{r}\hat{m}_{k+j}(\delta_{l,k} - \delta_{q,k}) + \sum_{|j|=1}^{r}\hat{m}_k(\delta_{l,k+j} - \delta_{q,k+j}) + \sum_{|j|=1}^{r}(\delta_{q,k}\delta_{l,k+j} + \delta_{q,k+j}\delta_{l,k})\right)\hat{\Pi}_{lq}^{++}. \tag{22}$$

Using Eq. (22) in Eq. (21), one can deduce

$$\begin{aligned}\sum_{k=1}^{L}\hat{\mathcal{P}}_k^{\dagger}\left(\hat{n}_{k,-} + \hat{n}_{k,+} - 2\hat{n}_{k,-}\hat{n}_{k,+}\right)\left[\hat{\Pi}_{lq}^{++}, \hat{\mathcal{P}}_k\right] =& \frac{\kappa}{2r}\hat{\mathcal{P}}_l^{\dagger}\left(\hat{n}_{l,-} + \hat{n}_{l,+} - 2\hat{n}_{l,-}\hat{n}_{l,+}\right)\sum_{|j|=1}^{r}\left(\hat{m}_{l+j} + \delta_{q,l+j}\right)\hat{\Pi}_{lq}^{++} \\ &+ \frac{\kappa}{2r}\hat{\mathcal{P}}_q^{\dagger}\left(\hat{n}_{q,-} + \hat{n}_{q,+} - 2\hat{n}_{q,-}\hat{n}_{q,+}\right)\sum_{|j|=1}^{r}\left(-\hat{m}_{q+j} + \delta_{l,q+j}\right)\hat{\Pi}_{lq}^{++} \\ &+ \frac{\kappa}{2r}\sum_{|j|=1}^{r}\left[\hat{\mathcal{P}}_{l-j}^{\dagger}\hat{m}_{l-j}\right]\hat{\Pi}_{lq}^{++} - \frac{\kappa}{2r}\sum_{|j|=1}^{r}\left[\hat{\mathcal{P}}_{q-j}^{\dagger}\hat{m}_{q-j}\right]\hat{\Pi}_{lq}^{++},\end{aligned} \tag{23}$$

$$\begin{aligned}\sum_{k=1}^{L}\left[\hat{\mathcal{P}}_k^{\dagger}, \hat{\Pi}_{lq}^{++}\right]\hat{\mathcal{P}}_k\left(\hat{n}_{k,-} + \hat{n}_{k,+} - 2\hat{n}_{k,-}\hat{n}_{k,+}\right) =& \frac{\kappa}{2r}\hat{\Pi}_{lq}^{++}\hat{\mathcal{P}}_l\left(\hat{n}_{l,-} + \hat{n}_{l,+} - 2\hat{n}_{l,-}\hat{n}_{l,+}\right)\sum_{|j|=1}^{r}\left(-\hat{m}_{l+j} + \delta_{q,l+j}\right) \\ &+ \frac{\kappa}{2r}\hat{\Pi}_{lq}^{++}\hat{\mathcal{P}}_q\left(\hat{n}_{q,-} + \hat{n}_{q,+} - 2\hat{n}_{q,-}\hat{n}_{q,+}\right)\sum_{|j|=1}^{r}\left(\hat{m}_{q+j} + \delta_{l,q+j}\right) \\ &- \frac{\kappa}{2r}\hat{\Pi}_{lq}^{++}\sum_{|j|=1}^{r}\left[\hat{\mathcal{P}}_{l-j}\hat{m}_{l-j}\right] + \frac{\kappa}{2r}\hat{\Pi}_{lq}^{++}\sum_{|j|=1}^{r}\left[\hat{\mathcal{P}}_{q-j}\hat{m}_{q-j}\right],\end{aligned} \tag{24}$$

$$\hat{\mathcal{P}}_q^{\dagger}\left[\hat{\Pi}_{lq}^{++}, \hat{\mathcal{P}}_q\right] = \frac{\kappa}{2r}\hat{\mathcal{P}}_q^{\dagger}\left(-\sum_{|j|=1}^{r}\hat{m}_{q+j} + (\hat{m}_q + 1)\sum_{|j|=1}^{r}\delta_{l,q+j}\right)\hat{\Pi}_{lq}^{++}, \tag{25}$$

$$\hat{\mathcal{P}}_l^{\dagger}\left[\hat{\Pi}_{lq}^{++}, \hat{\mathcal{P}}_l\right] = \frac{\kappa}{2r}\hat{\mathcal{P}}_l^{\dagger}\left(\sum_{|j|=1}^{r}\hat{m}_{l+j} + (-\hat{m}_l + 1)\sum_{|j|=1}^{r}\delta_{q,l+j}\right)\hat{\Pi}_{lq}^{++}, \tag{26}$$

where we have used the property of hard-core bosons $\hat{c}^\dagger_{l,\sigma}\hat{c}^\dagger_{l,\sigma} = \hat{c}_{l,\sigma}\hat{c}_{l,\sigma} = 0$. The first term in the RHS of Eq. (23) and (24) is further simplified as

$$
\begin{aligned}
\frac{\kappa}{2r}\hat{\Pi}^{++}_{lq}\hat{\mathcal{P}}_l\left(\hat{n}_{l,-}+\hat{n}_{l,+}-2\hat{n}_{l,-}\hat{n}_{l,+}\right)\sum_{|j|=1}^{r}\left(-\hat{m}_{l+j}+\delta_{q,l+j}\right) = &+\left(\frac{\kappa}{2r}\right)^2\left(\sum_{|j|=1}^{r}\hat{m}_{l+j}\right)^2\left(\hat{n}_{l,-}\right)\hat{\Pi}^{++}_{lq} \\
&-\left(\frac{\kappa}{2r}\right)^2\left(\hat{n}_{l,-}\right)\sum_{|j|=1}^{r}\left(\hat{m}_{l+j}\right)\hat{\Pi}^{++}_{lq}\sum_{|j|=1}^{r}\delta_{q,l+j} \\
&-\frac{\kappa}{2r}\hat{\mathcal{P}}_l\left(\hat{n}_{l,-}\right)\sum_{|j|=1}^{r}\left(\hat{m}_{l+j}\right)\hat{\Pi}^{++}_{lq},
\end{aligned} \tag{27}
$$

$$
\begin{aligned}
\frac{\kappa}{2r}\hat{\Pi}^{++}_{lq}\hat{\mathcal{P}}_q\left(\hat{n}_{q,-}+\hat{n}_{q,+}-2\hat{n}_{q,-}\hat{n}_{q,+}\right)\sum_{|j|=1}^{r}\left(\hat{m}_{q+j}+\delta_{l,q+j}\right) = &+\left(\frac{\kappa}{2r}\right)^2\left(\sum_{|j|=1}^{r}\hat{m}_{q+j}\right)^2\left(1-\hat{n}_{q,-}\right)\hat{\Pi}^{++}_{lq} \\
&+\left(\frac{\kappa}{2r}\right)^2\left(1-\hat{n}_{q,-}\right)\sum_{|j|=1}^{r}\left(\hat{m}_{q+j}\right)\hat{\Pi}^{++}_{lq}\sum_{|j|=1}^{r}\delta_{l,q+j} \\
&+\frac{\kappa}{2r}\hat{\mathcal{P}}_q\left(1-\hat{n}_{q,-}\right)\sum_{|j|=1}^{r}\left(\hat{m}_{q+j}\right)\hat{\Pi}^{++}_{lq}.
\end{aligned} \tag{28}
$$

The last two terms in Eq. (23) and (24) can be further simplified as

$$
\begin{aligned}
\frac{\kappa}{2r}\sum_{|j|=1}^{r}\left(\hat{\mathcal{P}}^\dagger_{l-j}\hat{m}_{l-j}\right)\hat{\Pi}^{++}_{lq} &- \frac{\kappa}{2r}\sum_{|j|=1}^{r}\left(\hat{\mathcal{P}}^\dagger_{q-j}\hat{m}_{q-j}\right)\hat{\Pi}^{++}_{lq} - \frac{\kappa}{2r}\hat{\Pi}^{++}_{lq}\sum_{|j|=1}^{r}\left(\hat{\mathcal{P}}_{l-j}\hat{m}_{l-j}\right) + \frac{\kappa}{2r}\hat{\Pi}^{++}_{lq}\sum_{|j|=1}^{r}\left(\hat{\mathcal{P}}_{q-j}\hat{m}_{q-j}\right) \\
&= -\frac{\kappa}{2r}\sum_{|j|=1}^{r}\left[\hat{\Pi}^{++}_{lq},\hat{\mathcal{P}}_{l-j}\hat{m}_{l-j}\right] + \frac{\kappa}{2r}\sum_{|j|=1}^{r}\left[\hat{\Pi}^{++}_{lq},\hat{\mathcal{P}}_{q-j}\hat{m}_{q-j}\right]
\end{aligned} \tag{29}
$$

and,

$$
\begin{aligned}
-\frac{\kappa}{2r}\sum_{|j|=1}^{r}\left[\hat{\Pi}^{++}_{lq},\hat{\mathcal{P}}_{l-j}\hat{m}_{l-j}\right] &+ \frac{\kappa}{2r}\sum_{|j|=1}^{r}\left[\hat{\Pi}^{++}_{lq},\hat{\mathcal{P}}_{q-j}\hat{m}_{q-j}\right] \\
&= \frac{\kappa}{2r}(-\hat{\mathcal{P}}_l-\hat{\mathcal{P}}_q)\hat{\Pi}^{++}_{lq} + \left(\frac{\kappa}{2r}\right)^2\sum_{|i|=1}^{r}(\hat{m}_{q+i}(1+\hat{m}_q)+\hat{m}_{l+i}(-1+\hat{m}_l))\hat{\Pi}^{++}_{lq} \\
&\quad + \left(\frac{\kappa}{2r}\right)^2 2(\hat{m}_l-\hat{m}_q)\hat{\Pi}^{++}_{lq} - \left(\frac{\kappa}{2r}\right)^2 2\hat{\Pi}^{++}_{lq}\sum_{|i|=1}^{r}\delta_{l,q+i}.
\end{aligned} \tag{30}
$$

Combining all these terms and substituting the expression of $\hat{\mathcal{P}}_k$ gives the final form of $\hat{\mathcal{C}}^{++}_{\mathcal{A}}$;

$$
\hat{\mathcal{C}}^{++}_{\mathcal{A}} = \left(\hat{\mathcal{C}}^{++}_{\mathcal{A}1} + \hat{\mathcal{C}}^{++}_{\mathcal{A}2}\sum_{|i|=1}^{r}\delta_{l,q+i}\right)\hat{\Pi}^{++}_{lq}, \tag{31}
$$

where

$$
\begin{aligned}
\hat{\mathcal{C}}^{++}_{\mathcal{A}1} =& -2 + \frac{\kappa}{r}\left(-1+\hat{M}_q(1-\hat{\rho}_q+\hat{m}_q)+\hat{M}_l(1-\hat{\rho}_l+\hat{m}_l)\right) \\
&+\left(\frac{\kappa}{2r}\right)^2\left(\hat{M}_q^2+\hat{M}_l^2+\hat{M}_q(1-\hat{\rho}_q+\hat{m}_q)+\hat{M}_l(1-\hat{\rho}_l+\hat{m}_l)+(2-\hat{\rho}_l+\hat{m}_l+\hat{\rho}_q-\hat{m}_q)\right), \\
\hat{\mathcal{C}}^{++}_{\mathcal{A}2} =& \frac{\kappa}{4r}\left(\hat{\rho}_q-\hat{m}_q-\hat{\rho}_l+\hat{m}_l\right)+\left(\frac{\kappa}{2r}\right)^2\left(\hat{M}_q-\hat{M}_l-2\right).
\end{aligned} \tag{32}
$$

Here, we have used $\hat{n}_{i,\sigma} = (\hat{\rho}_i + \sigma\hat{m}_i)/2$ and $\hat{M}_i = \sum_{|j|=1}^{r} \hat{m}_{i+j}$. The summation $\sum_{|j|=1}^{r} \delta_{l,q+j}$ gives a non-zero value only if $r \geq |q-l|$, which we always assume since we are in the mean-field approximation. Doing similar calculations for other values of $\sigma_1$ and $\sigma_2$ will give the general structure of $\hat{\mathcal{C}}_{\mathcal{A}}^{\sigma_1\sigma_2}$ as

$$\begin{aligned}\hat{\mathcal{C}}_{\mathcal{A}}^{\sigma_1\sigma_2} =& \Bigg[ -2 + \frac{\kappa}{r}\Big( -\sigma_1\sigma_2 + \sigma_2\hat{M}_q\left(1-(\hat{\rho}_q + \bar{\sigma_2}\hat{m}_q)\right) + \sigma_1\hat{M}_l\left(1-(\hat{\rho}_l + \bar{\sigma_1}\hat{m}_l)\right) + \frac{1}{4}\left(\hat{\rho}_q + \bar{\sigma_2}\hat{m}_q\right) - \frac{1}{4}\left(\hat{\rho}_l + \bar{\sigma_1}\hat{m}_l\right)\Big) \\ &+ \left(\frac{\kappa}{2r}\right)^2 \Big( -\sigma_1\bar{\sigma_2}\hat{M}_q^2 - \sigma_2\bar{\sigma_1}\hat{M}_l^2 + \sigma_1\hat{M}_q\left(2-(\hat{\rho}_q + \bar{\sigma_2}\hat{m}_q)\right) - \sigma_2\hat{M}_l\left(\hat{\rho}_l + \bar{\sigma_1}\hat{m}_l\right) - \left(\hat{\rho}_l + \bar{\sigma_1}\hat{m}_l\right) + \left(\hat{\rho}_q + \bar{\sigma_2}\hat{m}_q\right)\Big)\Bigg]\hat{\Pi}_{lq}^{\sigma_1\sigma_2}.\end{aligned} \tag{33}$$

## III. Homogeneous approximation to the dynamics of correlation functions

We use homogeneous approximation in the bulk as $\langle\hat{\rho}_i\rangle = \rho(t)$, $\langle\hat{m}_i\rangle = m(t)$, and $\langle M_i\rangle = \bar{M}$. We will discuss the homogeneous treatment of the each terms in the time evolution of the correlation functions in the following sections. We use Wick's theorem-like factorization for nonlinear terms as

$$\left\langle \hat{n}_{q,\sigma_1}\hat{\Pi}_{lq}^{\sigma_1\sigma_2}\right\rangle = \langle\hat{n}_{q,\sigma_1}\rangle\left\langle\hat{\Pi}_{lq}^{\sigma_1\sigma_2}\right\rangle + \left\langle\hat{\Pi}_{qq}^{\sigma_1\sigma_2}\right\rangle\left\langle\hat{\Pi}_{lq}^{\sigma_1\sigma_1}\right\rangle, \tag{34}$$

where the first term in the RHS corresponds to the mean-field level of factorization, while the second term related to the rate of spin-flips.

### A. Homogeneous approximation of $\left\langle\hat{\mathcal{C}}_H^{\sigma_1\sigma_2}\right\rangle$

We have

$$\begin{aligned}\left\langle\hat{\mathcal{C}}_H^{\sigma_1\sigma_2}\right\rangle =& \left\langle \hat{c}_{l,\sigma_1}^{\dagger}(1-2\hat{n}_{q,\sigma_2})\hat{c}_{q,\bar{\sigma}_2} - \hat{c}_{l,\bar{\sigma}_1}^{\dagger}(1-2\hat{n}_{l,\sigma_1})\hat{c}_{q,\sigma_2}\right\rangle \\ =& \left\langle\hat{\Pi}_{lq}^{\sigma_1\bar{\sigma_2}}\right\rangle - \left\langle\hat{\Pi}_{lq}^{\bar{\sigma_1}\sigma_2}\right\rangle - 2\left\langle\hat{\Pi}_{lq}^{\sigma_1\bar{\sigma_2}}\hat{n}_{q,\sigma_2}\right\rangle + 2\left\langle\hat{\Pi}_{lq}^{\bar{\sigma_1}\sigma_2}\hat{n}_{l,\sigma_1}\right\rangle.\end{aligned} \tag{35}$$

Doing the above mentioned factorization gives:

$$\begin{aligned}\left\langle\hat{\Pi}_{lq}^{\sigma_1\bar{\sigma_2}}\hat{n}_{q,\sigma_2}\right\rangle =& \left\langle\hat{\Pi}_{lq}^{\sigma_1\bar{\sigma_2}}\left(1-\hat{c}_{q,\sigma_2}\hat{c}_{q,\sigma_2}^{\dagger}\right)\right\rangle \\ =& \left\langle\hat{\Pi}_{lq}^{\sigma_1\bar{\sigma_2}}\right\rangle + \langle\hat{n}_{q,\sigma_2}\rangle\left\langle\hat{\Pi}_{lq}^{\sigma_1\bar{\sigma_2}}\right\rangle - \left\langle\hat{\Pi}_{lq}^{\sigma_1\sigma_2}\right\rangle\left\langle\hat{\Pi}_{qq}^{\sigma_2\bar{\sigma_2}}\right\rangle,\end{aligned} \tag{36}$$

$$\begin{aligned}\left\langle\hat{\Pi}_{lq}^{\bar{\sigma_1}\sigma_2}\hat{n}_{l,\sigma_1}\right\rangle =& \left\langle\hat{\Pi}_{lq}^{\bar{\sigma_1}\sigma_2}\left(1-\hat{c}_{l,\sigma_1}\hat{c}_{l,\sigma_1}^{\dagger}\right)\right\rangle \\ =& \left\langle\hat{\Pi}_{lq}^{\bar{\sigma_1}\sigma_2}\right\rangle + \langle\hat{n}_{l,\sigma_1}\rangle\left\langle\hat{\Pi}_{lq}^{\bar{\sigma_1}\sigma_2}\right\rangle - \left\langle\hat{\Pi}_{ll}^{\bar{\sigma_1}\sigma_1}\right\rangle\left\langle\hat{\Pi}_{lq}^{\sigma_1\sigma_2}\right\rangle,\end{aligned} \tag{37}$$

and substituting everything back gives

$$\left\langle\hat{\mathcal{C}}_H^{\sigma_1\sigma_2}\right\rangle = \left(-1-2\langle\hat{n}_{q,\sigma_2}\rangle\right)\left\langle\hat{\Pi}_{lq}^{\sigma_1\bar{\sigma_2}}\right\rangle + \left(1+2\langle\hat{n}_{l,\sigma_1}\rangle\right)\left\langle\hat{\Pi}_{lq}^{\bar{\sigma_1}\sigma_2}\right\rangle + 2\left(\left\langle\hat{\Pi}_{qq}^{\sigma_2\bar{\sigma_2}}\right\rangle - \left\langle\hat{\Pi}_{ll}^{\bar{\sigma_1}\sigma_1}\right\rangle\right)\left\langle\hat{\Pi}_{lq}^{\sigma_1\sigma_2}\right\rangle. \tag{38}$$

For the last term in Eq. (38), we assume homogeneity as $\left\langle\hat{\Pi}_{qq}^{\sigma_2\bar{\sigma_2}}\right\rangle = \Pi^{\sigma_2\bar{\sigma_2}}$, independent of lattice sites. This gives

$$\left\langle\hat{\Pi}_{qq}^{\sigma_2\bar{\sigma_2}}\right\rangle - \left\langle\hat{\Pi}_{ll}^{\bar{\sigma_1}\sigma_1}\right\rangle = \begin{cases} -if & \text{when} \quad \sigma_1,\sigma_2 = + \\ if & \text{when} \quad \sigma_1,\sigma_2 = - \\ 0 & \text{when} \quad \sigma_1 \neq \sigma_2, \end{cases} \tag{39}$$

where $f$ is the homogeneous spin-flip current defined in the main text.

### B. Homogeneous approximation of $\left\langle \hat{\mathcal{C}}_{\mathcal{M}}^{\sigma_1\sigma_2} \right\rangle$

We do the same factorization here,

$$\begin{aligned}\left\langle \hat{\mathcal{C}}_{\mathcal{M}}^{\sigma_1\sigma_2} \right\rangle =& \left\langle (\hat{n}_{q-\sigma_2,\sigma_2} - \hat{n}_{q+\sigma_2,\sigma_2} + \hat{n}_{l-\sigma_1,\sigma_1} - \hat{n}_{l+\sigma_1,\sigma_1} - 2)\,\hat{\Pi}_{lq}^{\sigma_1\sigma_2} \right\rangle \\ =& \left(\langle \hat{n}_{q-\sigma_2,\sigma_2}\rangle - \langle \hat{n}_{q+\sigma_2,\sigma_2}\rangle + \langle \hat{n}_{l-\sigma_1,\sigma_1}\rangle - \langle \hat{n}_{l+\sigma_1,\sigma_1}\rangle - 2\right) \left\langle \hat{\Pi}_{lq}^{\sigma_1\sigma_2} \right\rangle \\ &+ \left\langle \hat{\Pi}_{q-\sigma_2,q}^{\sigma_2\sigma_2} \right\rangle \left\langle \hat{\Pi}_{l,q-\sigma_2}^{\sigma_1\sigma_2} \right\rangle - \left\langle \hat{\Pi}_{q+\sigma_2,q}^{\sigma_2\sigma_2} \right\rangle \left\langle \hat{\Pi}_{l,q+\sigma_2}^{\sigma_1\sigma_2} \right\rangle + \left\langle \hat{\Pi}_{l-\sigma_1,q}^{\sigma_1\sigma_2} \right\rangle \left\langle \hat{\Pi}_{l,l-\sigma_1}^{\sigma_1\sigma_1} \right\rangle - \left\langle \hat{\Pi}_{l+\sigma_1,q}^{\sigma_1\sigma_2} \right\rangle \left\langle \hat{\Pi}_{l,l+\sigma_1}^{\sigma_1\sigma_1} \right\rangle,\end{aligned} \tag{40}$$

and in the homogeneous state, we get

$$\left\langle \hat{\mathcal{C}}_{\mathcal{M}}^{\sigma_1\sigma_2} \right\rangle = -\,2 \left\langle \hat{\Pi}_{lq}^{\sigma_1\sigma_2} \right\rangle. \tag{41}$$

The correlation functions depend on the distance between the lattice sites, and in this light, all other terms with opposite signs can be canceled.

### C. Homogeneous approximation of $\left\langle \hat{\mathcal{C}}_{\mathcal{A}}^{\sigma_1\sigma_2} \right\rangle$

We split $\hat{\mathcal{C}}_{\mathcal{A}}^{\sigma_1\sigma_2}$ as

$$\hat{\mathcal{C}}_{\mathcal{A}}^{\sigma_1\sigma_2} = -\,2\langle \hat{\Pi}_{lq}^{\sigma_1\sigma_2} \rangle + \frac{\kappa}{r}\hat{\mathcal{C}}_{\mathcal{A}(1)}^{\sigma_1\sigma_2} + \left(\frac{\kappa}{r}\right)^2 \hat{\mathcal{C}}_{\mathcal{A}(2)}^{\sigma_1\sigma_2}, \tag{42}$$

where $\hat{\mathcal{C}}_{\mathcal{A}(1)}^{\sigma_1\sigma_2}$ and $\hat{\mathcal{C}}_{\mathcal{A}(2)}^{\sigma_1\sigma_2}$ are the linear order and second order coefficients. Then,

$$\begin{aligned}\left\langle \hat{\mathcal{C}}_{\mathcal{A}(1)}^{\sigma_1\sigma_2} \right\rangle =& \left\langle \Big( -\sigma_1\sigma_2 + \sigma_2\hat{M}_q + \sigma_1\hat{M}_l + \big(-\sigma_2\hat{M}_q + \frac{1}{4}\big)\,(\hat{\rho}_q + \bar{\sigma_2}\hat{m}_q) + \big(-\sigma_1\hat{M}_l - \frac{1}{4}\big)\,(\hat{\rho}_l + \bar{\sigma_1}\hat{m}_l) \Big)\hat{\Pi}_{lq}^{\sigma_1\sigma_2} \right\rangle \\ =& \Big( -\sigma_1\sigma_2 + \sigma_2 \left\langle \hat{M}_q \right\rangle + \sigma_1 \left\langle \hat{M}_l \right\rangle \Big) \left\langle \hat{\Pi}_{lq}^{\sigma_1\sigma_2} \right\rangle \\ &+ \Big( \big(-\sigma_2 \left\langle \hat{M}_q \right\rangle + \frac{1}{4}\big)\,(\langle \hat{\rho}_q\rangle + \bar{\sigma_2}\,\langle \hat{m}_q\rangle) + \big(-\sigma_1 \left\langle \hat{M}_l \right\rangle - \frac{1}{4}\big)\,(\langle \hat{\rho}_l\rangle + \bar{\sigma_1}\,\langle \hat{m}_l\rangle) \Big) \left\langle \hat{\Pi}_{lq}^{\sigma_1\sigma_2} \right\rangle \\ &+ \big(-\sigma_2 \left\langle \hat{M}_q \right\rangle + \frac{1}{4}\big) \left\langle \hat{\Pi}_{qq}^{\bar{\sigma_2}\sigma_2} \right\rangle \left\langle \hat{\Pi}_{lq}^{\sigma_1\bar{\sigma_2}} \right\rangle + \big(-\sigma_1 \left\langle \hat{M}_l \right\rangle - \frac{1}{4}\big) \left\langle \hat{\Pi}_{lq}^{\bar{\sigma_1}\sigma_2} \right\rangle \left\langle \hat{\Pi}_{ll}^{\sigma_1\bar{\sigma_1}} \right\rangle,\end{aligned} \tag{43}$$

$$\begin{aligned}\left\langle \hat{\mathcal{C}}_{\mathcal{A}(2)}^{\sigma_1\sigma_2} \right\rangle =& \left\langle \Big( -\sigma_1\bar{\sigma_2}\hat{M}_q^2 - \sigma_2\bar{\sigma_1}\hat{M}_l^2 + 2\sigma_1\hat{M}_q + \Big(-\sigma_1\hat{M}_q + 1\Big)\,(\hat{\rho}_q + \bar{\sigma_2}\hat{m}_q) + \Big(-\sigma_2\hat{M}_l - 1\Big)\,(\hat{\rho}_l + \bar{\sigma_1}\hat{m}_l) \Big)\,\hat{\Pi}_{lq}^{\sigma_1\sigma_2} \right\rangle \\ =& \Big( -\sigma_1\bar{\sigma_2} \left\langle \hat{M}_q^2 \right\rangle - \sigma_2\bar{\sigma_1} \left\langle \hat{M}_l^2 \right\rangle + 2\sigma_1 \left\langle \hat{M}_q \right\rangle \Big) \left\langle \hat{\Pi}_{lq}^{\sigma_1\sigma_2} \right\rangle \\ &+ \Big( \Big(-\sigma_1 \left\langle \hat{M}_q \right\rangle + 1\Big)\,(\langle \hat{\rho}_q\rangle + \bar{\sigma_2}\,\langle \hat{m}_q\rangle) + \Big(-\sigma_2 \left\langle \hat{M}_l \right\rangle - 1\Big)\,(\langle \hat{\rho}_l\rangle + \bar{\sigma_1}\,\langle \hat{m}_l\rangle) \Big) \left\langle \hat{\Pi}_{lq}^{\sigma_1\sigma_2} \right\rangle \\ &+ \Big(-\sigma_1 \left\langle \hat{M}_q \right\rangle + 1\Big) \left\langle \hat{\Pi}_{qq}^{\bar{\sigma_2}\sigma_2} \right\rangle \left\langle \hat{\Pi}_{lq}^{\sigma_1\bar{\sigma_2}} \right\rangle + \Big(-\sigma_2 \left\langle \hat{M}_l \right\rangle - 1\Big) \left\langle \hat{\Pi}_{lq}^{\bar{\sigma_1}\sigma_2} \right\rangle \left\langle \hat{\Pi}_{ll}^{\sigma_1\bar{\sigma_1}} \right\rangle.\end{aligned} \tag{44}$$

After replacing expectation with homogeneous quantities, we get

$$\begin{aligned}\left\langle \hat{\mathcal{C}}_{\mathcal{A}(1)}^{\sigma_1\sigma_2} \right\rangle =& \left(-\sigma_1\sigma_2 + \sigma_2\bar{M} + \sigma_1\bar{M}\right) \left\langle \hat{\Pi}_{lq}^{\sigma_1\sigma_2} \right\rangle \\ &+ \Big( \big(-\sigma_1\bar{M} + \frac{1}{4}\big)\,(\rho + \bar{\sigma_2}m) + \big(-\sigma_2\bar{M} - \frac{1}{4}\big)\,(\rho + \bar{\sigma_1}m) \Big) \left\langle \hat{\Pi}_{lq}^{\sigma_1\sigma_2} \right\rangle \\ &+ \big(-\sigma_2\bar{M} + \frac{1}{4}\big) \left\langle \hat{\Pi}_{qq}^{\bar{\sigma_2}\sigma_2} \right\rangle \left\langle \hat{\Pi}_{lq}^{\sigma_1\bar{\sigma_2}} \right\rangle + \big(-\sigma_1\bar{M} - \frac{1}{4}\big) \left\langle \hat{\Pi}_{lq}^{\bar{\sigma_1}\sigma_2} \right\rangle \left\langle \hat{\Pi}_{ll}^{\sigma_1\bar{\sigma_1}} \right\rangle,\end{aligned} \tag{45}$$

$$\begin{aligned}\left\langle \hat{\mathcal{C}}_{\mathcal{A}(2)}^{\sigma_1\sigma_2} \right\rangle =& \Big( -2\sigma_1\bar{\sigma_2}\bar{M}^2 + 2\sigma_1\bar{M} \Big) \left\langle \hat{\Pi}_{lq}^{\sigma_1\sigma_2} \right\rangle \\ &+ \Big( \left(-\sigma_1\bar{M} + 1\right)(\rho + \bar{\sigma_2}m) + \left(-\sigma_2\bar{M} - 1\right)(\rho + \bar{\sigma_1}m) \Big) \left\langle \hat{\Pi}_{lq}^{\sigma_1\sigma_2} \right\rangle \\ &+ \Big( \left(-\sigma_1\bar{M} + 1\right) \left\langle \hat{\Pi}_{qq}^{\bar{\sigma_2}\sigma_2} \right\rangle \left\langle \hat{\Pi}_{lq}^{\sigma_1\bar{\sigma_2}} \right\rangle + \left(-\sigma_2\bar{M} - 1\right) \left\langle \hat{\Pi}_{lq}^{\bar{\sigma_1}\sigma_2} \right\rangle \left\langle \hat{\Pi}_{ll}^{\sigma_1\bar{\sigma_1}} \right\rangle \Big).\end{aligned} \tag{46}$$

### D. Final set of equations

Further, we write the homogeneous equation in compact matrix form. We start by using

$$\left\langle \hat{\Pi}_{qq}^{+-} \right\rangle = \frac{\epsilon - if}{2}, \qquad \left\langle \hat{\Pi}_{qq}^{-+} \right\rangle = \frac{\epsilon + if}{2}, \tag{47}$$

where $f = i\left\langle \hat{c}_{l,+}^{\dagger}\hat{c}_{l,-} - \hat{c}_{l,+}\hat{c}_{l,-}^{\dagger} \right\rangle$ and $\epsilon = \left\langle \hat{c}_{l,+}^{\dagger}\hat{c}_{l,-} + \hat{c}_{l,+}\hat{c}_{l,-}^{\dagger} \right\rangle$ are the homogeneous spin-flip current and the local energy, respectively. Then, we get the final set of coupled linear equations as

$$\frac{d}{dt}\begin{pmatrix} \left\langle \hat{\Pi}_{lq}^{++}(t) \right\rangle \\ \left\langle \hat{\Pi}_{lq}^{--}(t) \right\rangle \\ \left\langle \hat{\Pi}_{lq}^{+-}(t) \right\rangle \\ \left\langle \hat{\Pi}_{lq}^{-+}(t) \right\rangle \end{pmatrix} = \underbrace{\begin{pmatrix} c_+ & 0 & s_{1+} & s_{1-} \\ 0 & c_- & s_{2+} & s_{2-} \\ s_{3+} & s_{3-} & d_+ & 0 \\ s_{4+} & s_{4-} & 0 & d_- \end{pmatrix}}_{\mathcal{K}} \begin{pmatrix} \left\langle \hat{\Pi}_{lq}^{++}(t) \right\rangle \\ \left\langle \hat{\Pi}_{lq}^{--}(t) \right\rangle \\ \left\langle \hat{\Pi}_{lq}^{+-}(t) \right\rangle \\ \left\langle \hat{\Pi}_{lq}^{-+}(t) \right\rangle \end{pmatrix}, \tag{48}$$

where the elements of matrix $\mathcal{K}$ are given as follows:

$$\begin{aligned}
c_{\pm} =& \mp 2gf - \Gamma_{\mathcal{M}} - \Gamma_{\mathcal{A}}\left[1 + \frac{\kappa}{2r}\left(1 + 2\bar{M}\left(\mp 1 \pm \rho - m\right)\right) + \left(\frac{\kappa}{2r}\right)^2\left(-\bar{M}^2 + \bar{M}\left(\mp 1 \pm \rho - m\right)\right)\right], \\
d_{\pm} =& - \Gamma_{\mathcal{M}} - \Gamma_{\mathcal{A}}\left[1 + \frac{\kappa}{2r}\left(-1 + m\left(\pm\bar{M} - \frac{1}{4}\right)\right) + \left(\frac{\kappa}{2r}\right)^2\left(\bar{M}^2 \mp \bar{M} + m\left(\pm\bar{M} - 1\right)\right)\right], \\
s_{1\pm} =& \pm ig\left(1 - \rho - m\right) + \frac{\Gamma_{\mathcal{A}}}{4}\left(\frac{\kappa}{r}\left(-\bar{M} \pm \frac{1}{4}\right) + \left(\frac{\kappa}{2r}\right)^2\left(-\bar{M} \pm 1\right)\right)\left(\epsilon \pm if\right), \\
s_{2\pm} =& \pm ig\left(-1 + \rho - m\right) + \frac{\Gamma_{\mathcal{A}}}{4}\left(\frac{\kappa}{r}\left(\bar{M} \pm \frac{1}{4}\right) + \left(\frac{\kappa}{2r}\right)^2\left(\bar{M} \pm 1\right)\right)\left(\epsilon \pm if\right), \\
s_{3\pm} =& ig\left(\mp 1 \pm \rho - m\right) \pm \frac{\Gamma_{\mathcal{A}}}{4}\left(\frac{\kappa}{r}\left(\bar{M} - \frac{1}{4}\right) + \left(\frac{\kappa}{2r}\right)^2\left(\bar{M} - 1\right)\right)\left(\epsilon + if\right), \\
s_{4\pm} =& ig\left(\mp 1 \pm \rho - m\right) + \frac{\Gamma_{\mathcal{A}}}{4}\left(\frac{\kappa}{r}\left(-\bar{M} - \frac{1}{4}\right) + \left(\frac{\kappa}{2r}\right)^2\left(-\bar{M} - 1\right)\right)\left(\epsilon - if\right).
\end{aligned} \tag{49}$$

## IV. Solution for real part of the largest eigenvalue of $\mathcal{K}$

We start from the definition of the matrix $\mathcal{K}$:

$$\mathcal{K} = \begin{pmatrix} c_+ & 0 & s_{1+} & s_{1-} \\ 0 & c_- & s_{2+} & s_{2-} \\ s_{3+} & s_{3-} & d_+ & 0 \\ s_{4+} & s_{4-} & 0 & d_- \end{pmatrix}. \tag{50}$$

Neglecting contributions that are sub-leading in the interaction radius $r$, and using $\bar{M}/2r = m$ by setting maximum possible value for interaction radius $r = L/2$, the relevant constants take the form:

$$\begin{aligned}
c_{\pm} =& \mp 2gf - \Gamma_{\mathcal{M}} - \Gamma_{\mathcal{A}}\left[1 + \frac{\kappa}{2r} \mp \kappa m - m^2\left(\kappa^2 + 2\kappa\right)\right], \\
d_{\pm} =& - \Gamma_{\mathcal{M}} - \Gamma_{\mathcal{A}}\left[1 - \frac{\kappa}{2r} + m^2\left(\kappa^2 \pm \kappa\right)\right], \\
s_{1\pm} =& \frac{1}{2}\left[\pm ig\left(1 - 2m\right) - \Gamma_{\mathcal{A}}\kappa m\left(\epsilon \pm if\right)\right], \\
s_{2\pm} =& \frac{1}{2}\left[\mp ig\left(1 + 2m\right) + \Gamma_{\mathcal{A}}\kappa m\left(\epsilon \pm if\right)\right], \\
s_{3\pm} =& \frac{1}{2}\left[ig\left(\pm 1 - 2m\right) \pm \Gamma_{\mathcal{A}}\kappa m\left(\epsilon + if\right)\right], \\
s_{4\pm} =& \frac{1}{2}\left[ig\left(\mp 1 - 2m\right) \pm \Gamma_{\mathcal{A}}km\left(\epsilon - if\right)\right].
\end{aligned} \tag{51}$$

We can rewrite $\mathcal{K}$ as a block matrix:

$$\mathcal{K} = \begin{pmatrix} C & S \\ T & D \end{pmatrix} \quad \text{where} \quad C = \begin{pmatrix} c_+ & 0 \\ 0 & c_- \end{pmatrix}, D = \begin{pmatrix} d_+ & 0 \\ 0 & d_- \end{pmatrix}, S = \begin{pmatrix} s_{1+} & s_{1-} \\ s_{2+} & s_{2-} \end{pmatrix}, T = \begin{pmatrix} s_{3+} & s_{3-} \\ s_{4+} & s_{4-}. \end{pmatrix} \tag{52}$$

The eigenvalues can be calculated by solving the equation

$$\det\left[(C - \lambda I)(D - \lambda I) - ST\right] = 0, \tag{53}$$

or specifically by solving the equations

$$\begin{aligned} (\lambda - c_+)(\lambda - d_+) - \alpha =& 0, \quad \text{where} \quad \alpha = s_{1+}s_{3+} + s_{1-}s_{4+}, \\ (\lambda - c_-)(\lambda - d_-) - \beta =& 0, \quad \text{where} \quad \beta = s_{2+}s_{3-} + s_{2-}s_{4-}, \end{aligned} \tag{54}$$

which further gives the eigenvalues as

$$\begin{aligned} \lambda_{1,2} =& \frac{c_+ + d_+}{2} \pm \frac{1}{2}\sqrt{(c_+ - d_+)^2 + 4\alpha}, \\ \lambda_{3,4} =& \frac{c_- + d_-}{2} \pm \frac{1}{2}\sqrt{(c_- - d_-)^2 + 4\beta}. \end{aligned} \tag{55}$$

The real part of the largest eigenvalues can be either

$$\begin{aligned} \Re(\lambda_1) =& \frac{c_+ + d_+}{2} + \Re(\frac{1}{2}\sqrt{(c_+ - d_+)^2 + 4\alpha}) \quad \text{or} \\ \Re(\lambda_3) =& \frac{c_- + d_-}{2} + \frac{1}{2}\Re(\sqrt{(c_- - d_-)^2 + 4\beta}) \end{aligned} \tag{56}$$

because the terms $c$ and $d$ are real.

### A. Threshold condition where $\Re(\lambda_{max}) = 0$

Neutral stability occurs when $\Re(\lambda_1) = 0$, implying

$$(c_+ + d_+)^2 = (c_+ - d_+)^2 + 4\alpha. \tag{57}$$

Using the identity

$$(c_+ + d_+)^2 - (c_+ - d_+)^2 = 4c_+d_+, \tag{58}$$

we obtain the threshold equation

$$c_+d_+ = \alpha. \tag{59}$$

### B. Solving $c_+d_+ = \alpha$ to find the phase boundary, $\mathcal{K}_{cr}(g)$

Rewriting the expressions by substituting the elements of the matrix

$$\begin{aligned} c_+d_+ =& \left(-2gf - \Gamma_{\mathcal{M}} - \Gamma_{\mathcal{A}}\left[1 + \frac{\kappa}{2r} - \kappa m - m^2(\kappa^2 + 2\kappa)\right]\right)\left(-\Gamma_{\mathcal{M}} - \Gamma_{\mathcal{A}}\left[1 - \frac{\kappa}{2r} + m^2(\kappa^2 + \kappa)\right]\right), \\ \alpha =& -\frac{1}{2}\Gamma_{\mathcal{A}}\kappa m g f(1 - 2m) + \frac{\Gamma_{\mathcal{A}}^2\kappa^2 m^2}{4}(f^2 - \epsilon^2). \end{aligned} \tag{60}$$

Along with this, we have the following definitions from Eq. (10), (11) (13):

$$m^2 = \frac{1}{2\kappa_c^2}\frac{\kappa/\kappa_c - 1 - \Delta_g}{q - 1/\kappa_c}, \quad \Delta_g = 4g^2/\left(\Gamma_{\mathcal{A}}(\Gamma_{\mathcal{A}} + \Gamma_{\mathcal{M}})\right), \quad f = -\frac{4gm}{(\Gamma_{\mathcal{A}} + \Gamma_{\mathcal{M}})}, \quad \kappa_c(g) = (1 + \Delta_g)\kappa_0. \tag{61}$$

Here, the phase boundary between correlated and uncorrelated states is defined as the critical interaction strength as a function of the spin-flip amplitude, $\|_{cr}(g)$. To find this phase boundary, we have to solve Eq. (59) along with Eqs. (60) and (89).

### C. Phase boundary in the small $g$ limit

Let us assume that $\kappa(g)$ and $m(g)$ has a smooth even expansion,

$$\kappa(g) = \kappa_{cr}^0 + g^2\kappa_2 + g^4\kappa_4 + \mathcal{O}(g^6), \quad m(g) = m_0 + g^2 m_2 + g^4 m_4 + \mathcal{O}(g^4). \tag{62}$$

Now, we define the function

$$F(\kappa, m, g) = c_+ d_+ - \alpha, \tag{63}$$

which was assumed to have the following structure:

$$F(\kappa, m, g) = F_0 + g^2 F_2 + g^4 F_4. \tag{64}$$

At the phase boundary, $F$ vanishes. When $g = 0$, we get $F_0 = 0$. In the same way, for $g \neq 0$, the coefficient of higher order expansions should vanish in order to satisfy $F = 0$. The procedure is to substitute Eq. (62) in $F(\kappa, m, g)$ and set the coefficients of each order of $g$ to zero, and solve for $\kappa_0$, $\kappa_2$ and $\kappa_4$.

#### 1. Step 1: zeroth order terms in $g$

We set $g = 0$,

$$g = 0 \implies \Delta_g = 0, \kappa_c = \kappa_0, f = 0, \tag{65}$$

and this changes Eq. (64) as

$$F_0 = \left(-\Gamma_{\mathcal{M}} - \Gamma_{\mathcal{A}}\left[1 - \kappa m - m^2\left(\kappa^2 + 2\kappa\right)\right]\right)\left(-\Gamma_{\mathcal{M}} - \Gamma_{\mathcal{A}}\left[1 + m^2\left(\kappa^2 + \kappa\right)\right]\right) + \frac{\Gamma_{\mathcal{A}}^2 \kappa^2 m^2}{4}\epsilon^2. \tag{66}$$

Since we assume $|m| \ll 1$, we can neglect $m^2$ terms:

$$\left(-\Gamma_{\mathcal{M}} - \Gamma_{\mathcal{A}}\left[1 - \kappa m_0\right]\right)\left(-\Gamma_{\mathcal{M}} - \Gamma_{\mathcal{A}}\right) = 0, \tag{67}$$

where $m_0 = m(g = 0)$. The solution for $\kappa$ sets the phase boundary: $\kappa_{cr}(g = 0) = \kappa_{cr}^0(\Gamma_{\mathcal{A}}, \Gamma_{\mathcal{A}}, \kappa_0, q)$. For simplicity, we also use $\Gamma_{\mathcal{A}} = \Gamma_{\mathcal{M}}$. As we have the stability condition $q > 1/\kappa_c$, we can write $q = 1/\kappa_c + q_0$. Then, the zeroth order term of the solution reads:

$$\kappa_{cr}^0 = \frac{1}{3}\left(1 + \gamma^{-1/3} + \gamma^{1/3}\right)\kappa_0, \tag{68}$$

where $\gamma = (1 + 108q_0 + 6\sqrt{6}\sqrt{q_0(1 + 54q_0)})$. Interestingly, this result explains why the phase boundary shifts from the phase curve that separates disordered and flocking states. More precisely, as we make $q_0$ bigger, it moves the value of $\kappa_{cr}^0$ farther away from $\kappa_0$. A plot showing this shift is shown in Fig. 1.

#### 2. Step 2: second order terms in $g$

First, we determine $m_2$ using the equations Eq. (89) and (62):

$$m_2 = \frac{1}{4\kappa_0^2 m_0(1 - q/\kappa_0)}\left[\frac{\kappa_2}{\kappa_0} - \frac{4(2\kappa_{cr}^0/\kappa_0 - 1)}{\Gamma_{\mathcal{A}}(\Gamma_{\mathcal{A}} + \Gamma_{\mathcal{M}})}\right], \qquad \text{with} \quad m_0 = \sqrt{\frac{1}{2\kappa_0^2}\frac{\kappa_{cr}^0/\kappa_0 - 1}{q - 1/\kappa_c}}. \tag{69}$$

The structure of $F_2$ can be written as

$$F_2 = \frac{\partial F}{\partial \kappa}|_{\kappa=\kappa_0, g=0} \cdot \kappa_2 + \frac{\partial F}{\partial m}|_{m=m_0, g=0} \cdot m_2 + T_{exp}, \tag{70}$$

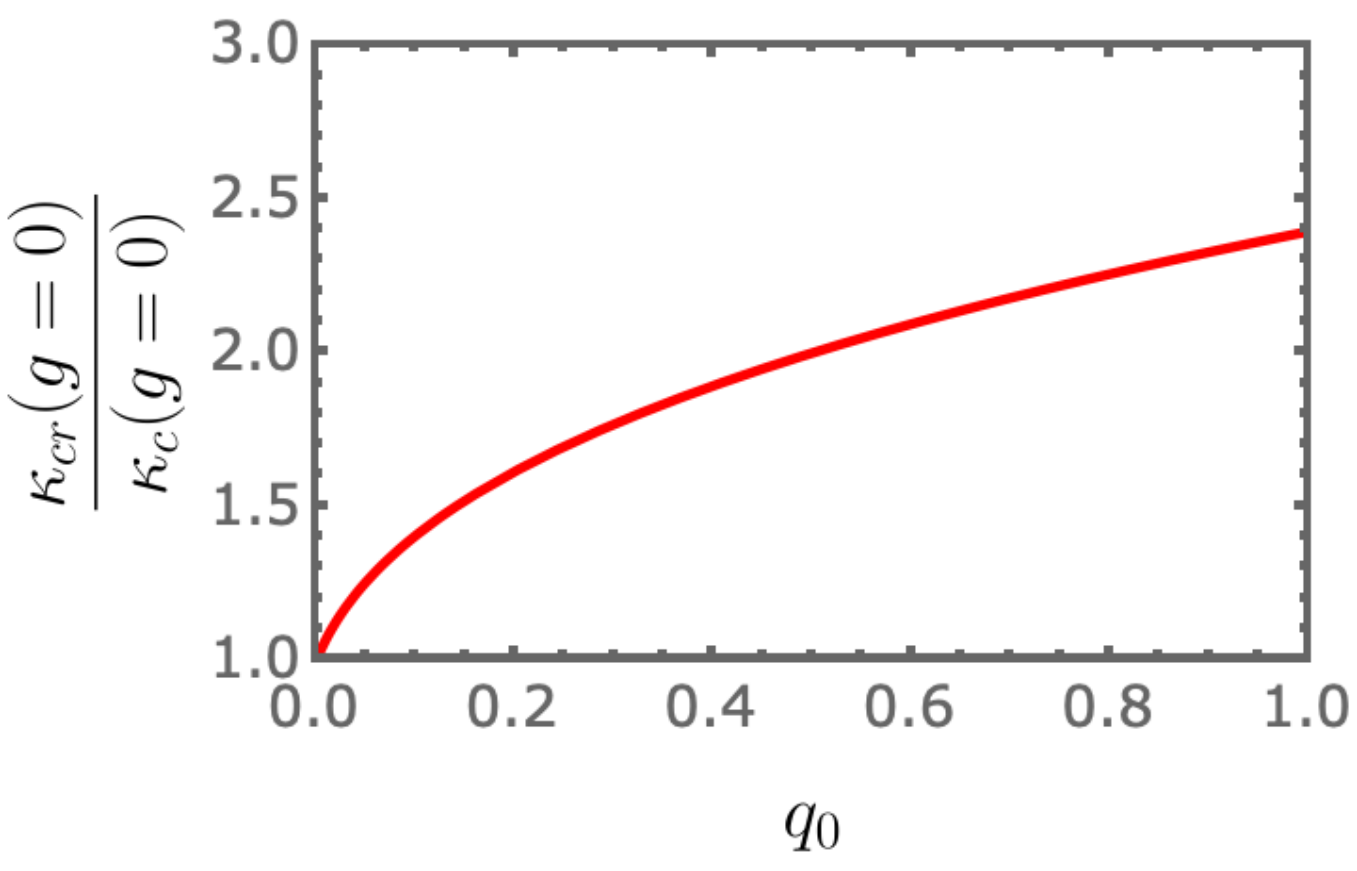


FIG. 1. $\kappa_{cr}(g=0)$ as a function of $q_0$

where the first and second terms in the RHS of the above equation show the change in $F_2$ as we change $\kappa(g)$ and $m(g)$, and the last term stands for terms that explicitly depend on $g$. Now, substituting the expression of $m_2$ gives the final expression of $\kappa_2$ as

$$\kappa_2 = -\frac{\frac{\partial F}{\partial m}|_{m=m_0,g=0}\left(\frac{1}{4\kappa_0^2 m_0(1-q/\kappa_0)}\left[-\frac{4(2\kappa_{cr}^0/\kappa_0-1)}{\Gamma_{\mathcal{A}}(\Gamma_{\mathcal{A}}+\Gamma_{\mathcal{M}})}\right]\right)+T_{exp}}{\frac{\partial F}{\partial \kappa}|_{\kappa=\kappa_0,g=0}+\frac{\partial F}{\partial m}|_{m=m_0,g=0}\left(\frac{1}{4\kappa_0^2 m_0(1-q/\kappa_0)}\left[\frac{1}{\kappa_0}\right]\right)}. \tag{71}$$

Performing the detailed calculation of each term leads to the expression

$$\kappa_2 = \frac{\frac{16\kappa_{cr}^0}{\Gamma_{\mathcal{A}}(\Gamma_{\mathcal{A}}+\Gamma_{\mathcal{M}})}+\frac{\Gamma_{\mathcal{A}}^2\kappa_0^2\epsilon^2}{2D_0}\frac{2\kappa_{cr}^0/\kappa_0-1}{\kappa_0(q-1/\kappa_0)}}{1-\frac{\Gamma_{\mathcal{A}}^2(\kappa_{cr}^0)^2\epsilon^2}{4D_0}\frac{1}{\kappa_0^2(q-1/\kappa_0)}}, \tag{72}$$

where $D_0 = \partial_\kappa(A_0B_0 + \frac{\Gamma_A^2\kappa^2m^2\epsilon^2}{4})|_{g=0}$, and $A_0, B_0$ are the brackets of $c_+d_+$ evaluated at $g = 0$.

Now, the derivative takes the form

$$D_0 = A_0^{'}B_0 + A_0B_0^{'} + \frac{\Gamma_A^2}{4}\partial_\kappa(\kappa^2m^2\epsilon^2). \tag{73}$$

To find the derivatives, we rewrite $m^2$ as

$$m^2 = \alpha(\kappa/\kappa_0 - 1), \quad \alpha = \frac{1}{2\kappa_c^2}\frac{1}{q-1/\kappa_0}, \tag{74}$$

which leads to

$$\partial m^2/\partial\kappa = \alpha/\kappa_0, \quad m\partial m/\partial\kappa = \alpha/2\kappa_0. \tag{75}$$

Further it gives,

$$\begin{aligned} A_0^{'} =& -\Gamma_A\left[1/2r - m - \kappa\partial m/\partial\kappa - (\alpha/\kappa_0)(\kappa^2+2\kappa) + m^2(2\kappa-2)\right] \\ B_0^{'} =& -\Gamma_A\left[-1/2r + (\alpha/\kappa_0)(\kappa^2+\kappa) + m^2(2\kappa+1)\right] \\ \frac{\Gamma_A^2}{4}\frac{\partial}{\partial\kappa}(\kappa^2m^2\epsilon^2) =& \frac{\Gamma_A^2\epsilon^2}{4}(2\kappa m^2 + \kappa^2\alpha/\kappa_0) \end{aligned} \tag{76}$$

We assume $m_0 \ll 1$, and $\kappa_{cr}^0 \sim \kappa_0$. Then

$$\begin{aligned} A_0 =& -\Gamma_M - \Gamma_A\left[1+1/2r\right] + \mathcal{O}(m_0^2), \\ B_0 =& -\Gamma_M - \Gamma_A\left[1-1/2r\right] + \mathcal{O}(m_0^2), \end{aligned} \tag{77}$$

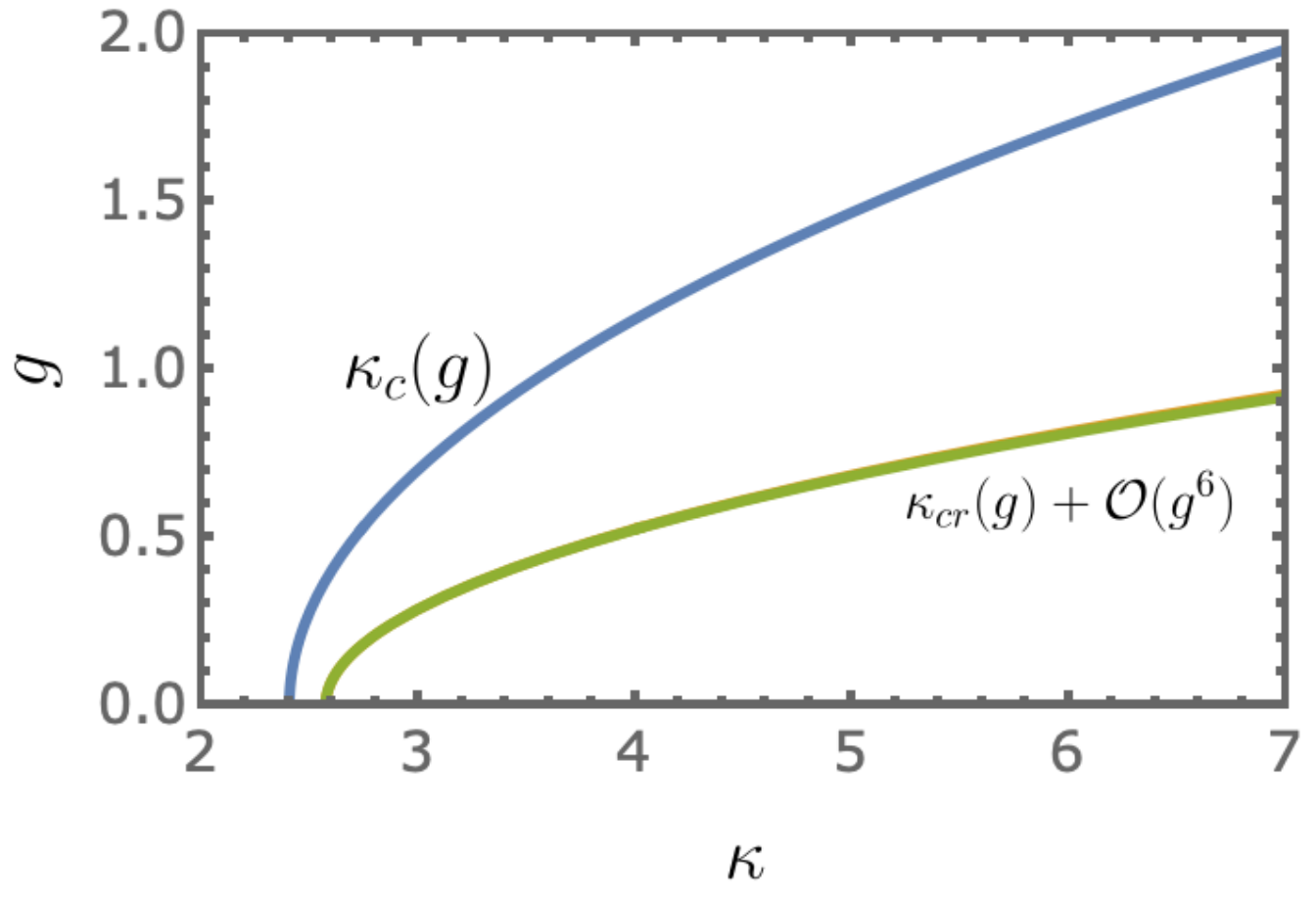


FIG. 2. Analytical phase diagram: Blue line represent the phase line separating disordered and flocking phase from the Eq. (13). The green lines represent analytical phase boundary separating correlated and uncorrelated motions, $\kappa_{cr}(g)$, from Eq. (82).

and their derivatives

$$\begin{aligned} A_0^{'} &= -\Gamma_A\left[1/2r - m_0 - \kappa_{cr}^0 \partial m/\partial\kappa + \mathcal{O}(m_0^2)\right], \quad \kappa_{cr}^0 \partial m/\partial\kappa \sim \alpha/(2\kappa_0 m_0), \\ B_0^{'} &= -\Gamma_A\left[-1/2r + \mathcal{O}(m_0^2)\right]. \end{aligned} \tag{78}$$

Since $A_0^{'} \sim \mathcal{O}(1/m_0)$, leading order term in $D_0$ is $D_0 \sim A_0^{'} B_0$. This leads to the result

$$\begin{aligned} D_0 =& \Gamma_A \alpha B_0 / 2\kappa_0 m_0 \\ =& \frac{\Gamma_A}{4\kappa_0^3 (q - 1/\kappa_0)} \frac{-\Gamma_M - \Gamma_A(1 - \kappa_0/2r)}{m_0}. \end{aligned} \tag{79}$$

Substituting this result back to Eq. (72), and putting $\epsilon = 0$ gives

$$\kappa_2 = \frac{16\kappa_{cr}^0}{\Gamma_{\mathcal{A}}(\Gamma_{\mathcal{A}} + \Gamma_{\mathcal{M}})} + \mathcal{O}(m_0^2) \tag{80}$$

### *3. Step 3: fourth order terms in g*

Following similar steps and making the same assumptions gives the fourth order term as

$$\kappa_4 = -\frac{m_0}{-\Gamma_{\mathcal{M}} - \Gamma_{\mathcal{A}}(1 - \kappa_0/2r)} \left[\frac{16(\kappa_{cr}^0)^2}{\Gamma_{\mathcal{A}}^2(\Gamma_{\mathcal{A}} + \Gamma_{\mathcal{M}})}\right] \tag{81}$$

### *4. Final expression $\kappa_{cr}(g)$*

Finally, the phase boundary is given by the following expression:

$$\kappa_{cr}(g) = \kappa_{cr}^0 + \frac{16\kappa_{cr}^0}{\Gamma_{\mathcal{A}}(\Gamma_{\mathcal{A}} + \Gamma_{\mathcal{M}})} g^2 + \left[\frac{16(\kappa_{cr}^0)^2 m_0}{\Gamma_{\mathcal{A}}^2(\Gamma_{\mathcal{A}} + \Gamma_{\mathcal{M}})^2}\right] g^4 + \mathcal{O}(g^6) \tag{82}$$

where we neglected the minor corrections that are proportional to $\epsilon^2$, to $\kappa_2$ and $\kappa_4$. In addition, the validity of this expression falls in a region where $g$ is small and $m_0 \ll 1$. A comparison of analytical phase boundaries separating disordered–flocking and correlated–uncorrelated regimes is shown in Fig. 2.

### V. Growth of the instability: eigenvalue and eigenvector analysis of $\mathcal{K}$

Here, we numerically analyses the growth of instabilities in correlators. Starting from the dynamics given by Eq. (48), one can see that the dynamics is set by matrix $\mathcal{K}$. The solution for $\mathbf{\Pi}$ can be written as follows:

$$\mathbf{\Pi}(t) = e^{\mathcal{K}t}\mathbf{\Pi}(0). \tag{83}$$

The correlators themselves are not eigenmodes, but the actual growing/decaying objects are the eigenvectors $v_n$ of matrix $\mathcal{K}$:

$$\mathcal{K}v_n = \lambda_n v_n, \tag{84}$$

and each mode evolves as $v_n \sim e^{\lambda_n t}$. The value of $\lambda_n$ and corresponding eigenvector $v_n$ determine which of the correlators grow in time and with what rate.

#### A. Calculating eigenvectors numerically for $m > 0$

We start by assuming majority of particles have spin (+), i.e., positive value for magnetization (Eq. (11)),

$$m = \sqrt{\frac{1}{2\kappa_c^2}\frac{\kappa/\kappa_c - 1 - \Delta_g}{q - 1/\kappa_c}}, \tag{85}$$

and interaction strength larger than critical interaction strength ($\kappa > \kappa_{cr}$). Then eigenvalues can be numerically calculated as follows (for $\kappa = 3.9$, $h = .3$, $\Gamma = 2$, and only real part is considered):

$$\begin{aligned} \lambda_1 =& 0.6, \\ \lambda_2 =& -3.1, \\ \lambda_3 =& -5, \\ \lambda_4 =& -6.1. \end{aligned} \tag{86}$$

The corresponding eigenvectors have the form (without the complex part)

$$v_{\lambda_1} = \begin{pmatrix} 1 \\ 0.002 \\ 0 \\ 0 \end{pmatrix}, \quad v_{\lambda_2} = \begin{pmatrix} 0 \\ 0 \\ 1 \\ -0.001 \end{pmatrix}, \quad v_{\lambda_3} = \begin{pmatrix} 0 \\ 0 \\ -0.02 \\ 1 \end{pmatrix}, \quad v_{\lambda_4} = \begin{pmatrix} 0 \\ 0 \\ 1 \\ .0004 \end{pmatrix}. \tag{87}$$

In the above numerical calculation, the largest eigen value is $\lambda_1$, which has a positive real part. Consequently, the first eigen mode is growing exponentially in time. Now, the corresponding eigenvector $v_{\lambda_1}$ has a approximate structure

$$v_{\lambda_1} \sim \begin{pmatrix} 1 \\ \varepsilon_0 \\ 0 \\ 0 \end{pmatrix}, \tag{88}$$

where $\varepsilon_0$ is small. This suggest that the correlator $\langle \Pi_{ij}^{++} \rangle$ is growing exponentially in time with largets growing rate, and mixing of $\langle \Pi_{ij}^{--} \rangle$ is considerably small. This instability growth could be the indication of a polarized state with superfludidity.

#### B. Calculating eigenvectors numerically for for $m < 0$

Doing the same analysis by keeping all parameters same and assuming the system posses a negative magnetization,

$$m = -\sqrt{\frac{1}{2\kappa_c^2}\frac{\kappa/\kappa_c - 1 - \Delta_g}{q - 1/\kappa_c}}, \tag{89}$$

the numerically calculated eigenvalues are

$$\begin{aligned} \lambda_1 =& -3.1, \\ \lambda_2 =& -6.1, \\ \lambda_3 =& -5, \\ \lambda_4 =& 0.6. \end{aligned} \tag{90}$$

The eigenvalue $\lambda_4$ is positive, and the corresponding eigenmode is growing exponentially in time, which has the structure

$$v_{\lambda_4} = \begin{pmatrix} 0.002 \\ 1 \\ 0 \\ 0 \end{pmatrix} \sim \begin{pmatrix} \varepsilon_0 \\ 1 \\ 0 \\ 0 \end{pmatrix}. \tag{91}$$

Consequently, the correlator $\langle \Pi_{ij}^{--} \rangle$ is growing exponentially in time with largest growing rate. This could indicate that the instability is leading to a polarized superfluid state.

### C. Detailed derivation of leading-order eigenvector mixing

We start with the assumption that real part of the largest eigenvalue of matrix $\mathcal{K}$ is positive but small:

$$\Re(\lambda_{max}) = \lambda_*, \quad \text{with} \quad \lambda_* \ll 1. \tag{92}$$

Now, we postulate the structure of the eigen vector corresponding to $\lambda_{max}$ as follows:

$$v_{\lambda_{max}} \sim \begin{pmatrix} 1 \\ \varepsilon_0 \\ 0 \\ 0 \end{pmatrix}, \tag{93}$$

and the motivation for this particular structure is from the numerical calculations. Our aim is to find the approximate scaling of $\varepsilon_0$, which will tell us about the mixing of correlation functions.

From the definition in Eq. (52), the eigenvector should satisfy:

$$\begin{pmatrix} C & S \\ T & D \end{pmatrix} \begin{pmatrix} x \\ y \end{pmatrix} = \lambda_* \begin{pmatrix} x \\ y \end{pmatrix}, \tag{94}$$

where $x = \begin{pmatrix} 1, & \varepsilon_0 \end{pmatrix}^T$ and $y = \begin{pmatrix} 0, & 0 \end{pmatrix}^T$. We eliminate $y$ by suing the identity

$$y = -(D - \lambda_* I)^{-1} T x. \tag{95}$$

This gives the effective equation for $\lambda_*$ as

$$\mathcal{K}_{\text{eff}}(\lambda) = C - S(D - \lambda_* I)^{-1} T, \quad \text{with} \quad \mathcal{K}_{\text{eff}} x = \lambda_* x. \tag{96}$$

Explicitly this reads

$$C - S(D - \lambda_* I)^{-1} T \begin{pmatrix} 1 \\ \varepsilon_0 \end{pmatrix} = \lambda_* \begin{pmatrix} 1 \\ \varepsilon_0 \end{pmatrix}. \tag{97}$$

In the LHS part, we can write the inverse operator $(D - \lambda_* I)^{-1}$ as

$$(D - \lambda_* I)^{-1} = \begin{pmatrix} \frac{1}{d_+ - \lambda_*} & 0 \\ 0 & \frac{1}{d_- - \lambda_*} \end{pmatrix}, \tag{98}$$

and the explicit form of LHS can be written as

$$C - S(D - \lambda_* I)^{-1} T = \begin{pmatrix} c_+ - \frac{s_{1+} s_{3+}}{d_+ - \lambda_*} - \frac{s_{1-} s_{4+}}{d_- - \lambda_*} & -\frac{s_{1+} s_{3-}}{d_+ - \lambda_*} - \frac{s_{1-} s_{4-}}{d_- - \lambda_*} \\ -\frac{s_{2+} s_{3+}}{d_+ - \lambda_*} - \frac{s_{2-} s_{4+}}{d_- - \lambda_*} & c_- - \frac{s_{2+} s_{3+}}{d_+ - \lambda_*} - \frac{s_{2-} s_{4-}}{d_- - \lambda_*} \end{pmatrix}. \tag{99}$$

Now, the expression for $\varepsilon_0$ can be written from Eq. (96):

$$\varepsilon_0 = \frac{\frac{s_{2+}s_{3+}}{d_+-\lambda_*} + \frac{s_{2-}s_{4+}}{d_--\lambda_*}}{c_- - \frac{s_{2+}s_{3+}}{d_+-\lambda_*} - \frac{s_{2-}s_{4-}}{d_--\lambda_*} - \lambda_*} \tag{100}$$

Using the explicit expressions of matrix elements in Eq. (51), one can deduce $\varepsilon_0$ in the orders of $m$ as

$$\varepsilon_0 = \frac{g^2}{2(2\Gamma+\lambda_*)^2} - mg^2\frac{(\kappa\Gamma - 2\lambda_* - 4\Gamma)}{2(2\Gamma+\lambda_*)^3} + \mathcal{O}(m^2), \tag{101}$$

or in the leading order

$$\varepsilon_0 = \frac{g^2}{2(2\Gamma+\lambda_*)^2} + \mathcal{O}(mg^2). \tag{102}$$

It is clear from the expression that the value of $\varepsilon_0$ always very small, $\varepsilon_0 \ll 1$. In the flocking phase, $g$ is taken as small and that keeps the value of $\varepsilon_0$ small. The effect of interaction strength $\kappa$ and magnetization $m$ enters through $\lambda_* = \lambda_*(g, \kappa, m, \Gamma)$. If the interaction strength is sufficiently strong, then $\lambda_*$ takes a large value and $\varepsilon_0$ still takes a small down.


[1] R. Khasseh, S. Wald, R. Moessner, C. A. Weber, and M. Heyl, Active quantum flocks, Phys. Rev. Lett. **135**, 248302 (2025).

[2] A. P. Solon and J. Tailleur, Flocking with discrete symmetry: The two-dimensional active ising model, Physical Review E **92**, 10.1103/physreve.92.042119 (2015).

 

%

\end{document}